\documentclass[showpacs,amssymb,floatfix]{revtex4}[14pt]
\usepackage{graphicx}
\usepackage{dcolumn}
\usepackage{graphics}
\usepackage{epstopdf}
\usepackage{pdfpages}
\usepackage{bm}
\begin{document}
\title{Near-bound states in the radiation continuum in circular array of dielectric rods}
\author{Evgeny N. Bulgakov$^{1,2}$ and Almas F. Sadreev$^1$}
\affiliation{$^1$ Kirensky Institute of Physics, Federal Research Center KSC SB RAS, 660036
Krasnoyarsk, Russia\\
$^2$Siberian State University of Science and Technology,
Krasnoyarsk 660037, Russia}
\date{\today}
\begin{abstract}
We consider E polarized bound states in the radiation continuum
(BICs)  in circular periodical arrays of $N$ infinitely long
dielectric rods. We find that each true BIC which occurs in an
infinite linear array has its counterpart in the circular array as
a near-BIC  with extremely large quality factor. We argue
analytically as well as numerically that the quality factor of the
symmetry protected near-BICs diverges as $e^{\lambda N}$ where
$\lambda$ is a material parameter dependent on the radius and the
refraction index of the rods. By tuning of the radius of rods we
also find numerically non-symmetry protected near-BICs. These
near-BICs are localized with exponential accuracy outside the
circular array but fill the whole inner space of the array
carrying orbital angular momentum.
\end{abstract}
 \maketitle
\section{Introduction}
Recently confined electromagnetic modes above the light line,
bound states in the continuum (BICs) were shown to exist in
periodic arrays of  long dielectric rods
\cite{Shipman0,Shipman,Marinica,Ndangali2010,Hsu Nature,Weimann,
Wei,Bo Zhen,Yang,PRA2014,Hu&Lu,Foley,Song,Zou,Yuan,Z
Wang,Yuan&Lu,Sadrieva}.
\begin{figure}[ht]
\includegraphics[width=10cm,clip=]{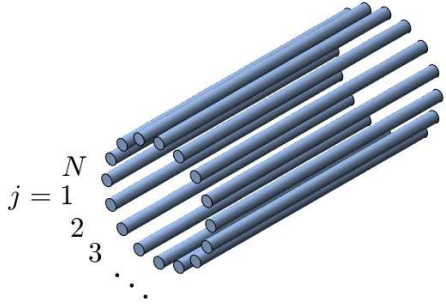}
\caption{(Color online) $N$ infinitely long circular dielectric
rods with radius $a$ are stacked parallel each other on a surface
of cylinder with the radius $R$ which is measured in terms of the
distance between centers of the nearest rods. } \label{fig1}
\end{figure}
Among numerous types of BICs it is worthy to emphasize the BICs
which can propagate either cross to rods \cite{PRA2014,Yuan&Lu} or
along the axis of periodicity of the system \cite{Wei}. In
practice there are no infinite arrays of rods. In the array of a
finite number of rods the BICs become quasi-BICs with finite
Q-factor which diverges as $N^2$ or even as $N^3$
\cite{Polishuk,B&MOE}. These results are in agreement with the
theorem on the absence of BICs in the bounded domain which is
complement of an unbounded domain \cite{Colton,Silveirinha}. The
BICs can appear only in unbounded domain. Physically occurrence of
the BICs in the infinite array of rods is a result of periodicity
of the array that quantizes the radiation continua in the form of
diffraction continua \cite{PRA2014,PRA2017}. Then if the frequency
is below the cut-off of the second diffraction continuum the BIC
is embedded into the first diffraction continuum. Note that the
second diffraction continuum is also important providing a bound
space for the BIC \cite{PRA2014}.

Moreover the scattering of acoustic waves and water surface waves
by array of cylinders was extensively studied in series of papers
\cite{Linton&Evans1990,Maniar}. The remarkable case of a circular
array of cylinders was considered in Refs.
\cite{Ludwig,Paknys,Evans&Porter1997,Fikioris}. Numerical results
of strong confinement of light in a circular array of dielectric
pillars \cite{Sieutat} and symmetry protected quasi-BICs with
exponentially high quality factor in the circular array of
dielectric nanorods \cite{Lu&Liu} were reported recently. We
reexamine these results for E polarized symmetry protected BICs in
the circular array of $N$ infinitely long cylindrical dielectric
rods and demonstrate the patterns of quasi-BICs with extremely
large quality factors. Following \cite{Evans&Porter1997} we define
such BICs as near-BICs.

We present mathematical arguments in favor of exponentially large
quality factors of the symmetry protected near-BICs in the
circular array of dielectric rods. In addition we find numerically
the non-symmetry protected near-BICs by tuning the rod radius. In
contrast to the symmetry protected near-BICs they fill the
internal space of the circular array. The circular array of rods
support non-symmetry protected near-BICs with orbital angular
momentum (OAM). Finally, we demonstrate the counterparts of the
BICs in the linear array embedded into two and three diffraction
continua which fill only a part of the inner space of circular
array. The diffraction continua for the linear chain is given by
plane waves
\begin{equation}\label{psizn}
E_z(x,y)=e^{iq_{y,p}y+i(q_p+q_x)x}
\end{equation}
where
\begin{equation}\label{diffr cont}
q_{y,p}=\sqrt{k_0^2-(q_x+q_p)^2}, q_p=\frac{2\pi p}{h}
\end{equation}
and $p=0, \pm 1, \pm 2, \ldots$ enumerates the diffraction
continua for the periodical chain of rods with the period $h$.

\section{Scattering theory for circular array of cylinders}
\begin{figure}[ht]
\includegraphics[width=6cm,clip=]{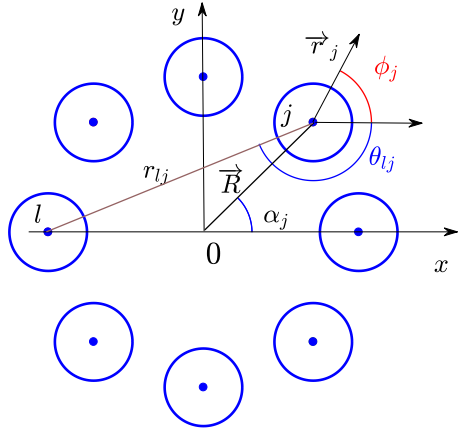}
\caption{Plan view of circular periodical array of rods.} \label{fig2}
\end{figure}

Following Ref. \cite{Yasumoto} (see quite similar procedure
described in ref. \cite{Linton&Evans1990} for the Neumann boundary
conditions at the surfaces of rigid cylinders) we present the
general E polarized solution at the vicinity of the j-th rod for
the electric field directed along the rods as follows
\begin{equation}\label{Ez}
    E_z(r_j,\phi_j)=E_{inc}+\sum_m[a_m(j)J_m(k_0r_j)+b_m(j)H_m^{(1)}(k_0r_j)]e^{im\phi_j},
\end{equation}
where the first term presents the incident wave from a point-like
source placed at the center of circular array $x=0, y=0$ as
sketched in Figs. \ref{fig2} and \ref{fig3}
\begin{equation}\label{inc}
E_{inc}=H_n^{(1)}(k_0r)e^{in\phi},
\end{equation}
the second term is a contribution of the other rods and the field
emanating from the $j$-th rod. $\phi_j$ and $r_j$ are the polar
coordinates of the radius-vector $\mathbf{r}_j$ in the local
coordinate systems of the $j$-th rod as shown in Fig. \ref{fig2}.
We introduce substitutions
\begin{equation}\label{ab}
    a_m(j)=\widetilde{a}_m(j)e^{-im\alpha_j}, ~~b_m(j)=\widetilde{b}_m(j)e^{-im\alpha_j}
\end{equation}
where $\alpha_j=\frac{2\pi(j-1)}{N}$ (see Fig. \ref{fig2}) and use
the Graf formula \cite{Yasumoto}:
\begin{equation}\label{Graf}
    H_m^{(1)}(k_0r_j(P))e^{im\phi_j}=\sum_{m'}e^{i(m-m')\theta_{jl}}H_{m'-m}^{(1)}(k_0r_{jl})
    J_{m'}(k_0r_l(P))e^{i\phi_l(P)}
\end{equation}
where definitions of angles and distances are shown in Fig.
\ref{fig2}. That allows us to write the following relations
between the amplitudes $\widetilde{a}_m$ and $\widetilde{b}_m$:
\begin{equation}\label{graf}
\widetilde{a}_m(j)=\sum_{l\neq
j}^N\sum_{m'}\widetilde{b}_{m'}(l)
\exp[im\alpha_j-im'\alpha_l-i(m-m')\theta_{lj})]H_{m-m'}^{(1)}(k_0r_{lj}).
\end{equation}

Periodicity of the circular array of rods allows us to write
\begin{eqnarray}\label{Bloch}
&\widetilde{a}_{mn}(j)=\widetilde{a}_{mn}(1)e^{ik_n(j-1)},&\nonumber\\
&\widetilde{b}_{mn}(j)=\widetilde{b}_{mn}(1)e^{ik_n(j-1)},
k_n=2\pi n/N, n=0,1,2,\ldots, N-1&
\end{eqnarray}
where $k_n$ is the Bloch number.
In particular for $j=1$ we have from Eqs. (\ref{graf}) and (\ref{Bloch})
\begin{equation}\label{rod1}
\widetilde{a}_{mn}(1)=\sum_{l=2}^N\sum_{m'}\widetilde{b}_{m'n}(1)
\exp[ik_n(l-1)-i(m-m')\theta_{l1}-im'\alpha_l]H_{m-m'}^{(1)}(k_0r_{lj}).
\end{equation}
Finally, we close Eq. (\ref{rod1}) which relates incident
amplitude $\widetilde{a}_{mn}(1)$ with emanating amplitude
$\widetilde{b}_{mn}(1)$ by equation
\begin{equation}\label{S}
    \widetilde{b}_{mn}(1)=S_m[\widetilde{a}_{mn}(1)+\psi_{inc,m}^n],
\end{equation}
where $S_m$ is the diagonal component of the S-matrix of a
circular dielectric rod
\begin{equation}
\label{eq3}
S_m=\frac{\sqrt{\epsilon}J^{'}_{m}(qa)J_m(k_0a)-J^{'}_{m}(k_0a)J_m(qa)}
{H^{(1)'}_{m}(k_0a)J_m(qa)-\sqrt{\epsilon}J^{'}_{m}(qa)H^{(1)}_m(k_0a)},
\end{equation}
$q=\sqrt{\epsilon}k_0$, and $\epsilon$ is the permittivity of the
rod of radius $a$. Substituting Eq. (\ref{S}) into Eq.
(\ref{rod1}) we can formulate the basic equation
\begin{equation}\label{LS}
    \sum_{m'}L_{mm'}^n\widetilde{b}_{m',n}(1)=\psi_{inc,m}^n
\end{equation}
where the matrix elements are given by
\begin{equation}\label{Lmm}
    L_{mm'}^n=-S_m\sum_{l=2}^N\exp[ik_n(l-1)+i(m'-m)\theta_{l1}-im'\alpha_l]H_{m-m'}^{(1)}(k_0r_{l1})
    +\delta_{mm'}
\end{equation}
and according to Eq. (\ref{inc})
\begin{equation}\label{psiinc}
\psi_{inc,m}^n=S_mH_{n-m}^{(1)}(k_0R).
\end{equation}

\section{Far-field zone solution}

Outside the rods for $r_j>a$ the solution with the Bloch wave number $k_n$ takes
the following form \cite{Yasumoto}
\begin{equation}\label{Ezr>a}
    E_{z,n}=\sum_m\sum_{j=1}^N\widetilde{b}_{mn}(1)\exp[ik_n(j-1)+im(\phi_j-\alpha_j)]H_m^{(1)}(k_0r_j)
    e^{im\phi_j}.
\end{equation}
\begin{figure}[ht]
\includegraphics[width=6cm,clip=]{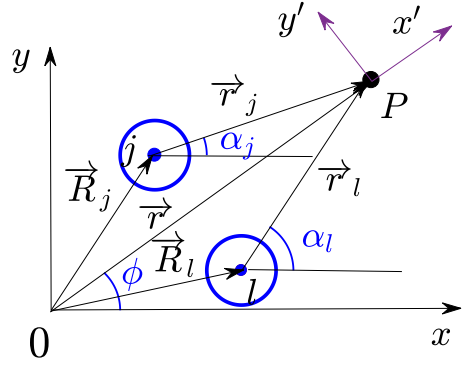}
\caption{The point $P$ in the far-field zone} \label{fig3}
\end{figure}
Now we can write the electric field in the far-field zone $r_j\gg
R$ at the point $P$ shown in Fig. \ref{fig3} by the use of the
asymptotic form of the Hankel function
\begin{equation}\label{asymp}
    H_m^{(1)}(x)\approx\sqrt{\frac{2}{\pi x}}\exp[-i\pi(2m+1)/4+ix].
\end{equation}
We have $\overrightarrow{r}=\overrightarrow{R}_j+\overrightarrow{r}_j$
where all three radius-vectors are shown in Fig. \ref{fig3}. For
$r\gg R$ we can approximate
\begin{equation}\label{Ra}
    \sqrt{\frac{1}{k_0r_j}}\exp(ik_0r_j)\approx \sqrt{\frac{1}{k_0r}}\exp[ik_0r-ik_0R\cos(\alpha_j-\phi))].
\end{equation}
Therefore in the far-field zone the electric field can be
approximated as follows
\begin{equation}\label{Ezfar}
    E_{z,n}\approx\sqrt{\frac{2}{k_0r}}e^{ik_0r-i\pi/4}\sum_m(-i)^m\widetilde{b}_{mn}
    \sum_{j=1}^N\exp[ik_n(j-1)+im(\phi-\alpha_j)-ik_0R\cos(\alpha_j-\phi)].
\end{equation}

Now we apply to this equation a mathematical observation of
exponential convergence of sums \cite{Trefethen}
\begin{equation}\label{IN}
    I_N=\frac{1}{N}\sum_{k=1}^Nu(e^{2\pi ik/N}).
    \end{equation}
The sum is converged as follows
\begin{equation}\label{conver}
    |I_N-I|\leq \frac{max(u)}{s^N-1},
\end{equation}
for some $s>1$ and
\begin{equation}\label{max}
    I=\frac{1}{2\pi}\int_0^{2\pi}d\theta u(e^{i\theta})
\end{equation}
where $u(\theta)=u(\theta+2\pi)$ is a periodical analytical
function. Therefore we can write Eq. (\ref{Ezfar}) as follows
\begin{equation}\label{Ezcont}
E_{z,n}=
Ne^{in\phi}\sum_m(-i)^m\widetilde{b}_{mn}\sqrt{\frac{2}{k_0r}}e^{ik_0r-i\pi/4}
\frac{1}{2\pi}\int_0^{2\pi}d\tau\exp[-i\tau(m-n)-ik_0R\cos\tau]+O(e^{-\lambda
N}).
\end{equation}
Using the identity for the Bessel functions
\begin{equation}\label{bessel}
    J_n(z)=\frac{i^{-n}}{\pi}\int_0^{\pi}e^{iz\cos\tau}\cos(n\tau)d\tau
\end{equation}
we have for the electric field (\ref{Ezcont}) in the far-zone
\begin{equation}\label{EzP}
E_z=E_{inc}+Ne^{in\phi}\sqrt{\frac{2}{k_0r}}e^{ik_0r-i\pi/4}
(-i)^n\sum_m\widetilde{b}_{mn}J_{n-m}(k_0R)+O(e^{-\lambda N}).
\end{equation}
\section{The solution inside the circular array of rods}
Now we consider the solution inside the circular array. We have
for the electric field
\begin{equation}\label{Ezminside}
    E_{z,n}=\sum_m\widetilde{b}_{mn}\sum_{j=1}^Ne^{i(n-m)\alpha_j}H_m^{(1)}(k_0r_j)e^{im\phi_j}.
\end{equation}
Let us use the Graf formula in order to transfer the solution at
local position defined by $\overrightarrow{r}_j=(r_j,\phi_j)$ to
the solution in the global system of coordinates defined by the
radius-vector $\overrightarrow{r}=(r,\phi)$. We have
\begin{equation}\label{graf1}
H_m^{(1)}(k_0r_j)e^{im\phi_j}=\sum_{m'}e^{i(m-m')\alpha_j}H^{(1)}_{m'-m}(k_0R) J_{m'}(k_0r)e^{im'\phi}.
\end{equation}
Substitution of this equation into Eq. (\ref{Ezminside}) gives
\begin{equation}\label{Ezminside1}
E_{z,n}=\sum_m\widetilde{b}_{mn}\sum_{m'}\sum_{j=1}^N
e^{i(k_n-k_{m'})(j-1)}H_{m'-m}^{(1)}(k_0R)
J_{m'}(k_0r)e^{im'\phi}.
\end{equation}
Due to the equality
\begin{equation}\label{DD}
    \sum_{j=1}^Ne^{i2\pi(j-1)(m-m')/N}=N\delta(m-m'-qN)
\end{equation}
where $q$ is an integer,
we can simplify Eq. (\ref{Ezminside1}) as follows
\begin{equation}\label{Ezinside2}
E_{z,n}=
N\sum_{m,q}\widetilde{b}_{mn}H_{n+qN-m}^{(1)}(k_0R)J_{n+qN}(k_0r)e^{i(n+qN)\phi}.
\end{equation}
\section{Near-BICs}
For the infinite periodical arrays if a source is switched off,
there are exceptional cases with selected real eigenfrequencies
embedded into the radiation continuum as was briefly reviewed in
the Introduction. These exceptional cases define BICs which are
localized in the vicinity of the  arrays
\cite{Shipman0,Shipman,Marinica,Ndangali2010,Hsu Nature,Weimann,
Wei,Bo Zhen,Yang,PRA2014,Hu&Lu,Foley,Song,Zou,Yuan,Z
Wang,Yuan&Lu,Sadrieva}. Let us consider what happens to these BICs
if to roll up a finite array into a circle. For the solution could
vanish in the far-field zone we have to imply that all
$\widetilde{b}_{mn}=0$ \cite{Silveirinha}. That in turn requires
that there is no electric field in the near-zone too according to
Eq. (\ref{Ezminside}), i.e., the solution equals zero everywhere.
However, if only exponential smallness is required in the
far-field zone,  for $N \gg 1$ we can imply a softer condition for
the amplitudes
\begin{equation}\label{BIC}
    \sum_m \widetilde{b}_{mn}J_{n-m}(k_0R)=0,
\end{equation}
as it is seen from Eq. (\ref{EzP}). Therefore, the solution turns
to zero in the far-field zone with exponential accuracy. While in
the near-field zone the solution given by Eq. (\ref{Ezinside2}) is
finite to be referred  to as a near-BIC.

Now we show that this formulation of the near-BICs in the circular
arrays is consistent with source-free Eq. (\ref{LS})
\begin{equation}\label{BICLS}
   \widehat{L}\Psi_c=0,
\end{equation}
which has no solution for real frequencies.  However, by
analytical continuation of the frequency
$k_0$ into the complex plane $\rm{Re}(k_0)+i\rm{Im}(k_0)$ we can
find a solution of Eq. (\ref{BICLS})  which has an extremely small
imaginary part $-\rm{Im}(k_0)\sim e^{-\lambda N}$. Numerics show
that the corresponding eigenmode $\Psi_c$ satisfies Eq.
(\ref{BIC}).
\subsection{The symmetry protected BICs}
We start a consideration with the solution which is odd relative
to $y'\rightarrow -y'$ in local system of coordinates as sketched
in Fig. \ref{fig3}. For $k_n=0$ (standing waves) we can take all
amplitudes $\widetilde{a}_m(j)$ and $\widetilde{b}_m(j)$ as
independent of the site index $j$. Hence we can rewrite the
solution (\ref{Ez}) in the vicinity of of the rods as follows
\begin{eqnarray}\label{Ezsin}
    &E_{z,even}=\sum_{m=0}^{\infty}[\widetilde{a}_{2m+1}J_{2m+1}(k_0r_j)+
    \widetilde{b}_{2m+1}(j)H_{2m+1}^{(1)}(k_0r_j)]\sin[(2m+1)(\phi_j-\alpha_j)],&\\
&E_{z,odd}=\sum_{m=1}^{\infty}[\widetilde{a}_{2m}(j)J_{2m}(k_0r_j)+
    \widetilde{b}_{2m}(j)H_{2m}^{(1)}(k_0r_j)]\sin[2m(\phi_j-\alpha_j)].&
\end{eqnarray}
For the odd solution we imply the following equalities
\begin{eqnarray}\label{bodd}
&\widetilde{b}_{2m+1}=\widetilde{b}_{-2m-1}&\nonumber\\
&\widetilde{b}_{2m}=-\widetilde{b}_{-2m},&
\end{eqnarray}
then Eq. (\ref{BIC}) is fulfilled. For the directional array of
infinite number of rods the most trivial symmetry protected BIC is
the solution which is odd relative to each rod in the array
direction. The solution of the symmetry protected BIC is presented
in Refs. \cite{PRA2014,Hu&Lu}. Then the coupling of the BIC with
the first diffraction continuum obviously equals zero.

\begin{figure}[ht]
\includegraphics[width=0.35\linewidth,clip=]{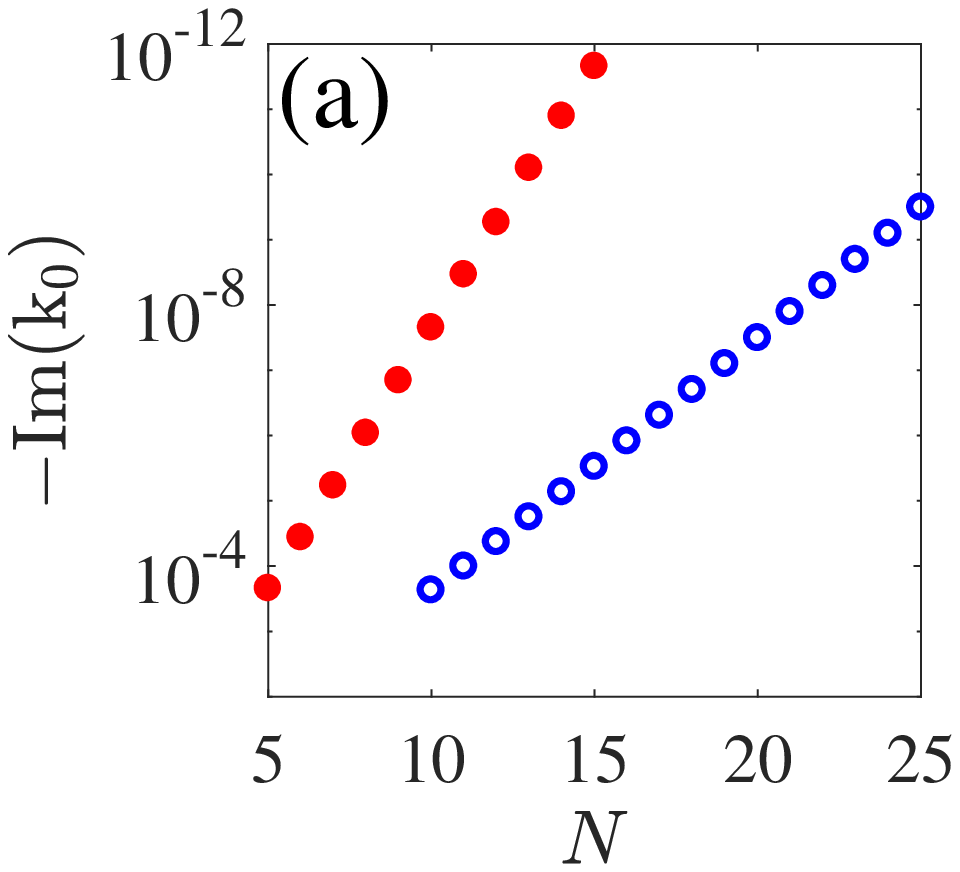}
\includegraphics[width=0.35\linewidth,clip=]{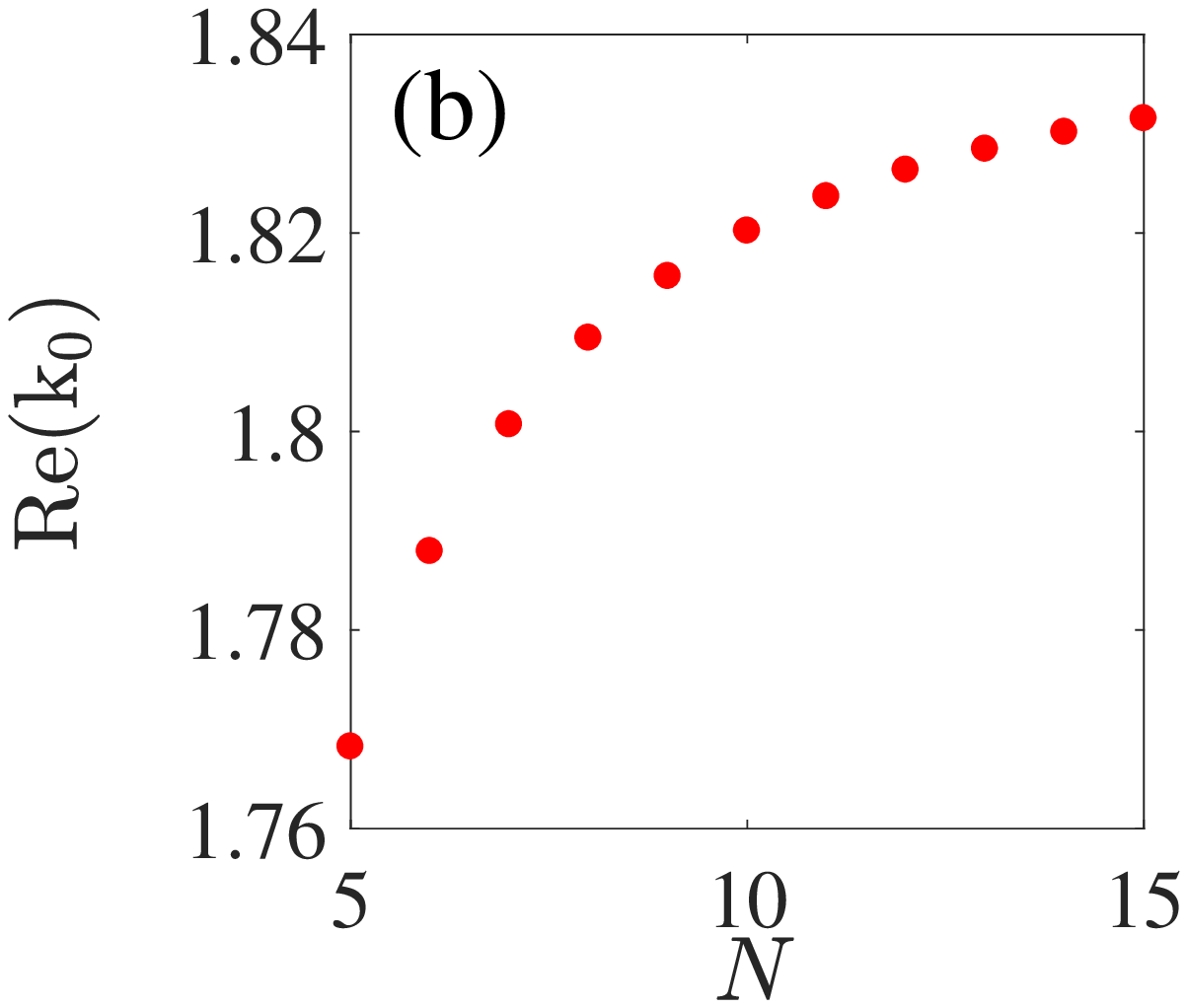}
\includegraphics[width=0.35\linewidth,clip=]{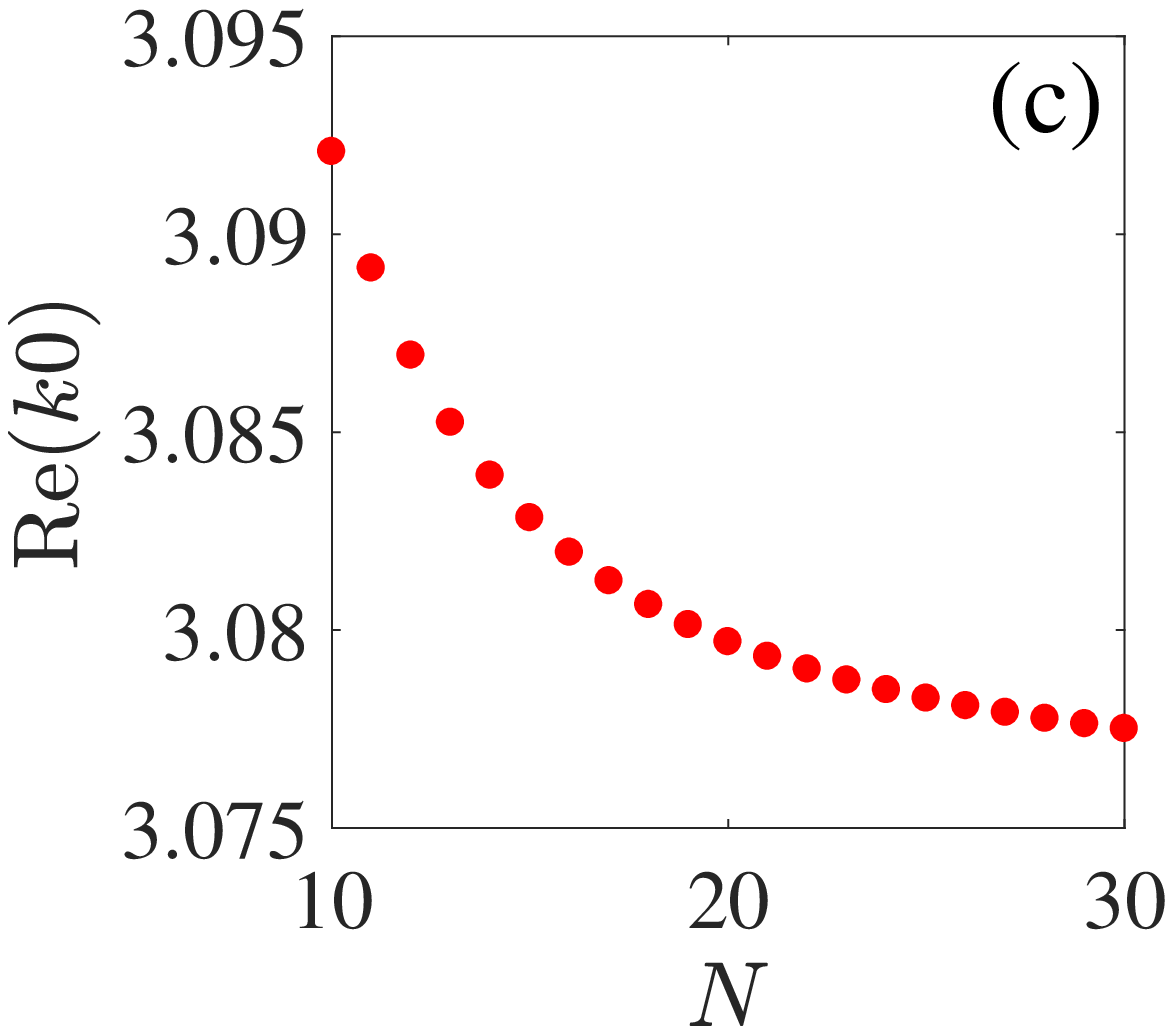}
\caption{The pole of the inverse of matrix (\ref{Lmm})
exponentially small in imaginary part
 versus the number of rods
for the symmetry protected near-BIC shown below in Figs.
\ref{fig5} and \ref{fig6} with parameters $\epsilon=15, a=0.44$.
(a) Imaginary part and (b) and (c) real part of the pole.}
\label{fig4}
\end{figure}

Below we hold in numerics two  parameters of dielectric rods
fixed: the permittivity $\epsilon=15$ (silicon rods)and the period
$h=2\pi R/N=1$ (the angular distance between centers of the rods).
The frequency $k_0$ is measured in terms of $c/h$ where $c$ is the
light velocity. In Fig. \ref{fig4} we show as the pole of the
matrix $\widehat{L}$ behaves with growth of the number of
rods for different types of the symmetry protected near-BICs. One
can see that the real part of this complex eigenvalue limits to
the frequency of the symmetry protected true BIC in the infinite array of rods.

\begin{figure}[ht]
\includegraphics[width=0.35\linewidth,clip=]{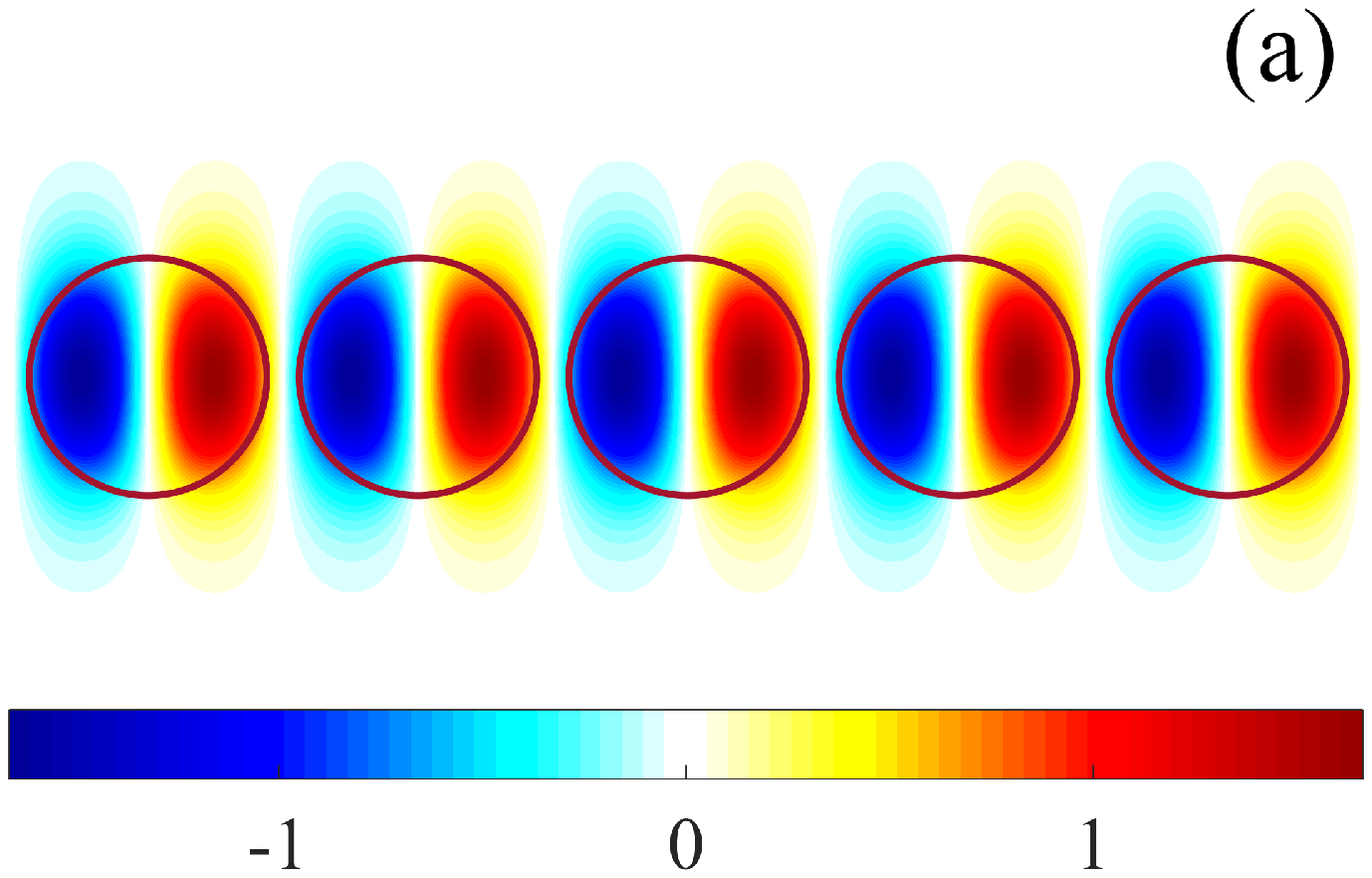}
\includegraphics[width=0.45\linewidth,clip=]{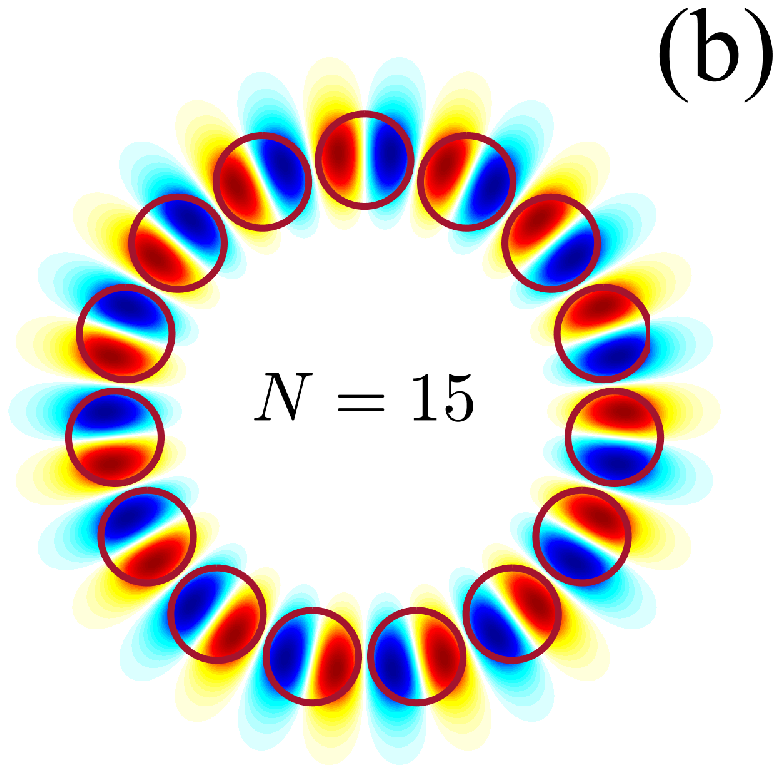}
\includegraphics[width=0.55\linewidth,clip=]{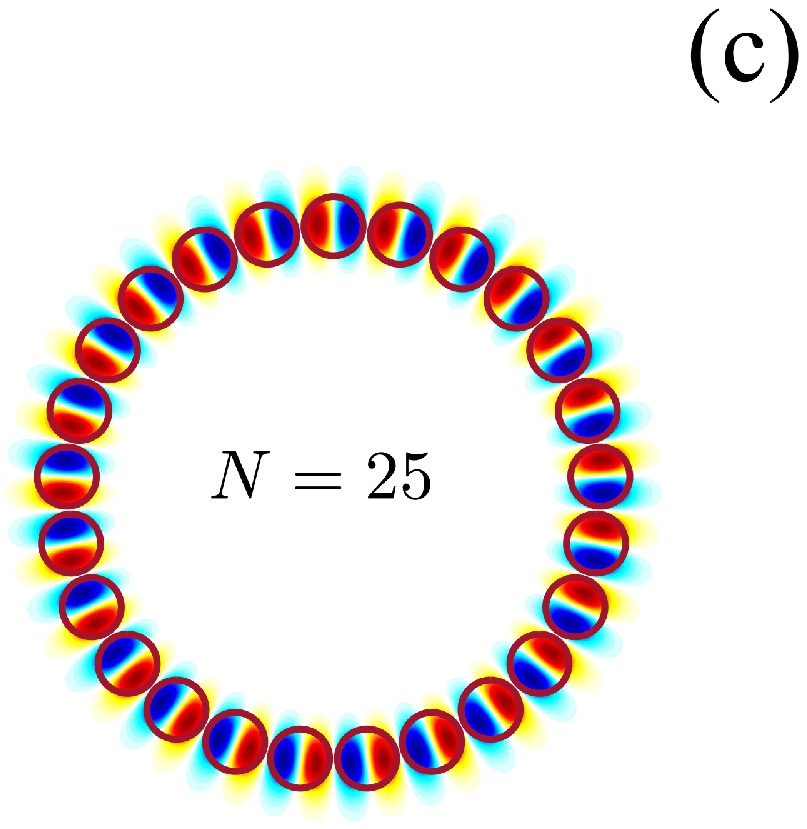}
\includegraphics[width=0.4\linewidth,clip=]{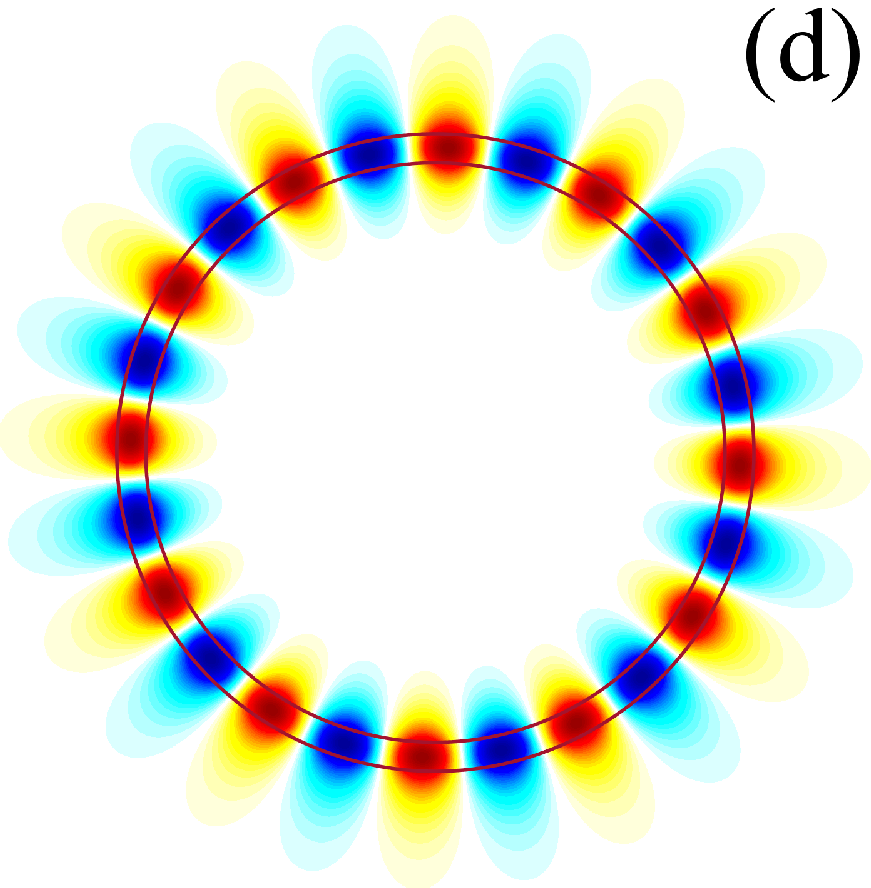}
\caption{Pattern of the true symmetry protected BIC in (a) linear
array with the frequency $k_0=1.8412$ and its counterparts in
circular array of 15 rods with the frequency $k_0=1.8315$ and
$Q=2\cdot 10^{11}$ (b) and of 25 rods with the frequency
$k_0=1.837$ and $Q=2\cdot 10^{20}$
 for $\epsilon=15, a=0.44$.
(d) Whispering gallery mode with angular momentum $m=12$ and
frequency $k_0=0.45304, Q=1.1\cdot 10^7$.} \label{fig5}
\end{figure}
\begin{figure}[ht]
\includegraphics[width=0.3\linewidth,clip=]{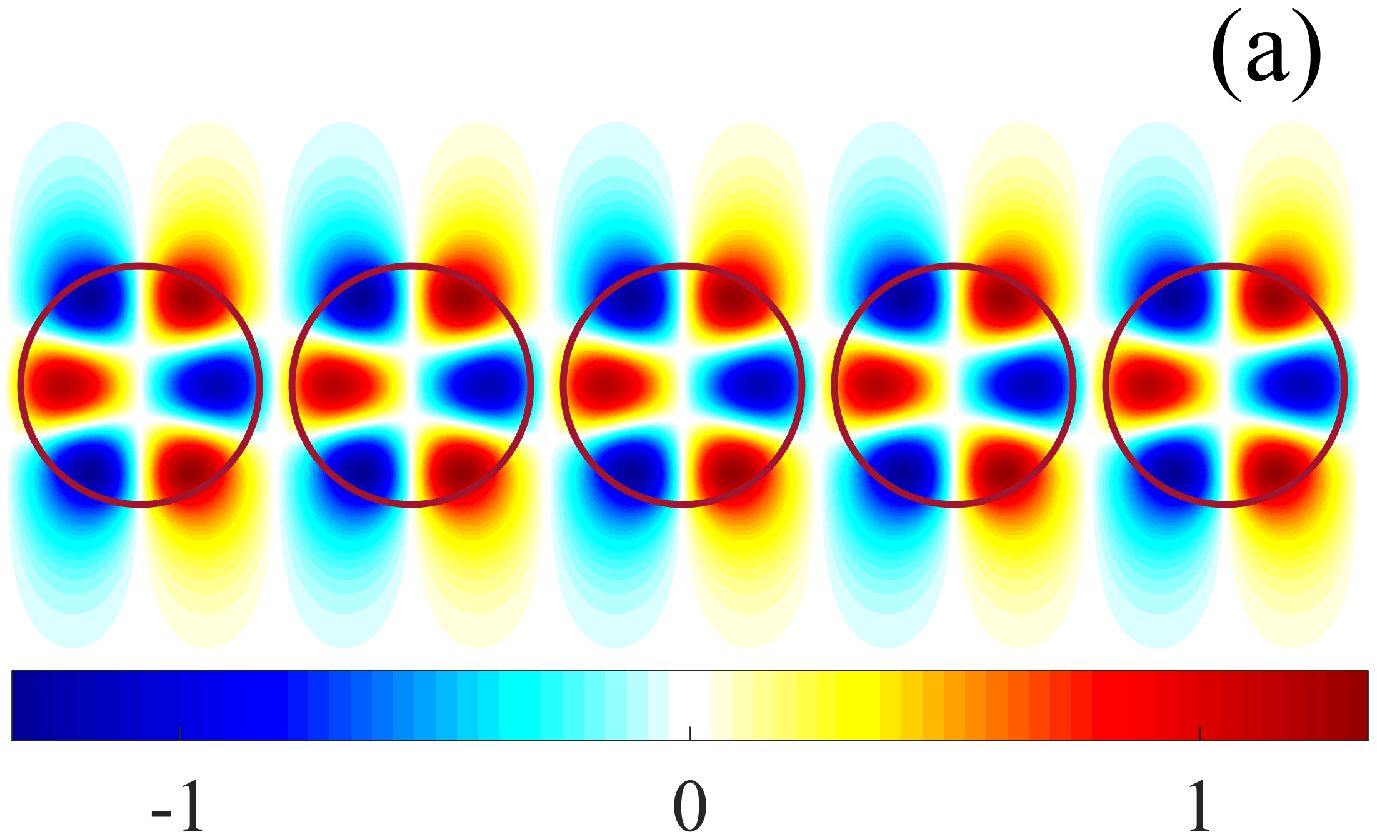}
\includegraphics[width=0.3\linewidth,clip=]{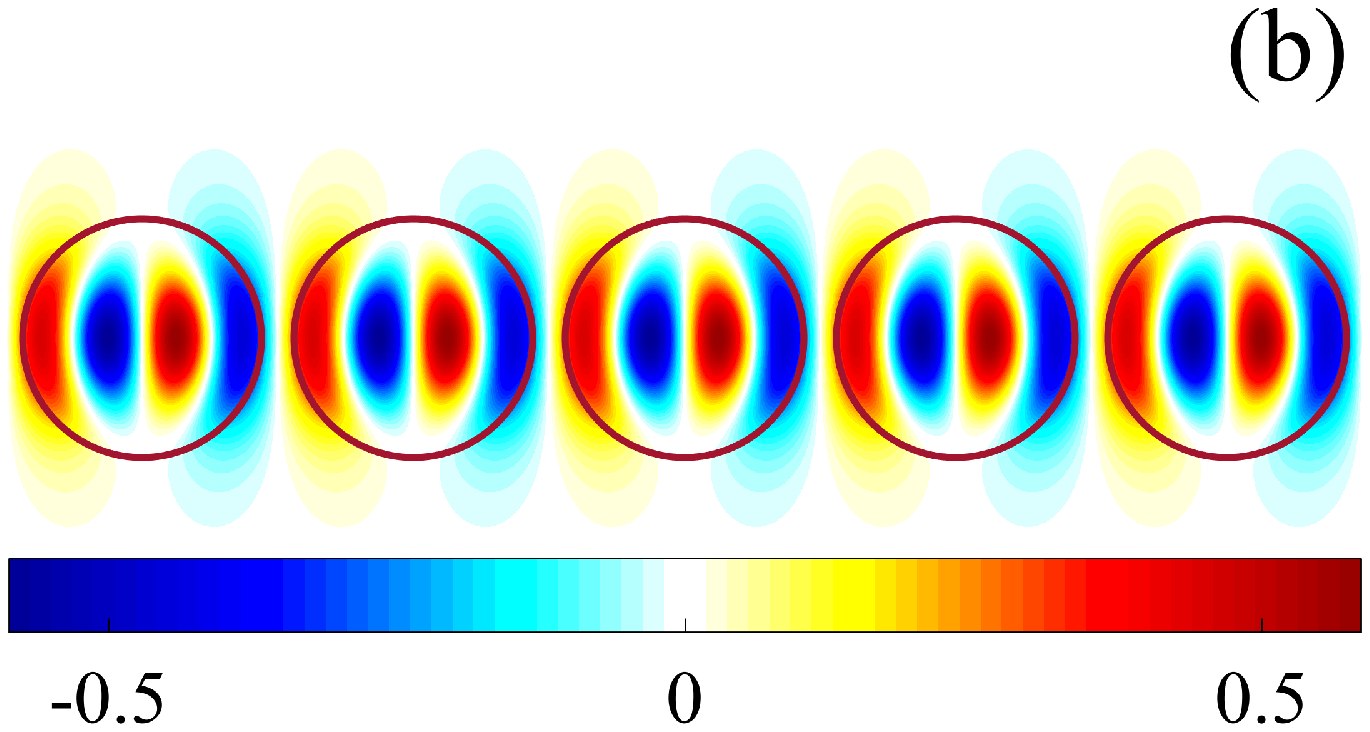}
\includegraphics[width=0.45\linewidth,clip=]{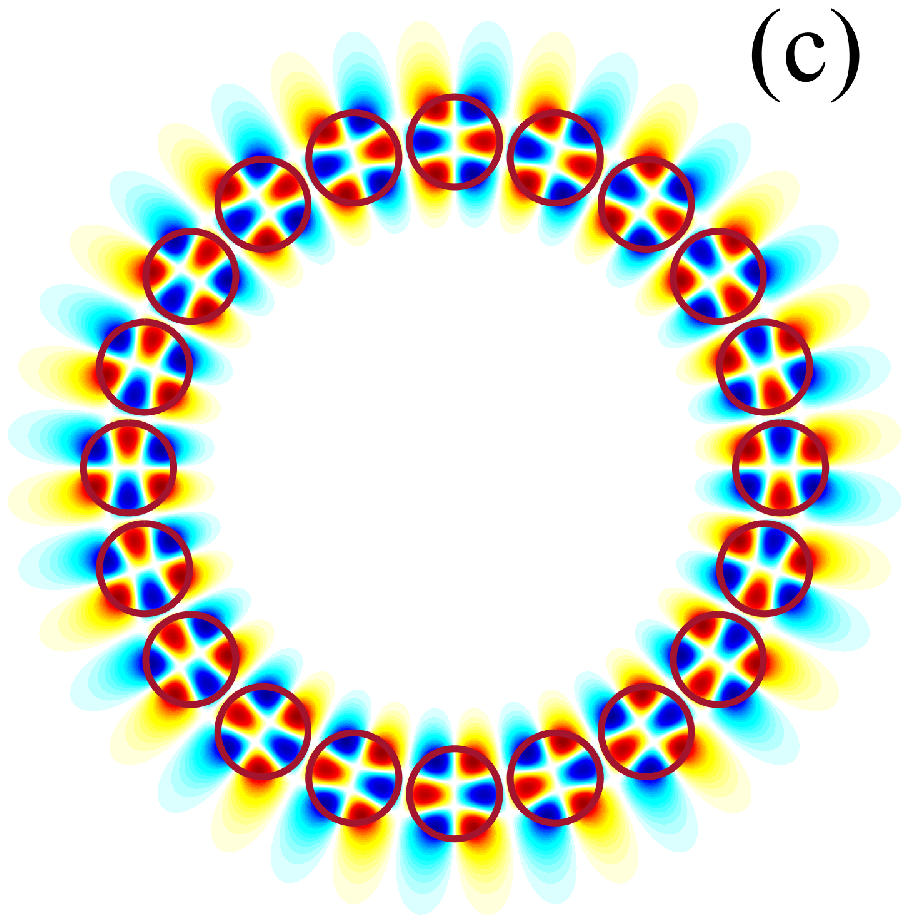}
\includegraphics[width=0.45\linewidth,clip=]{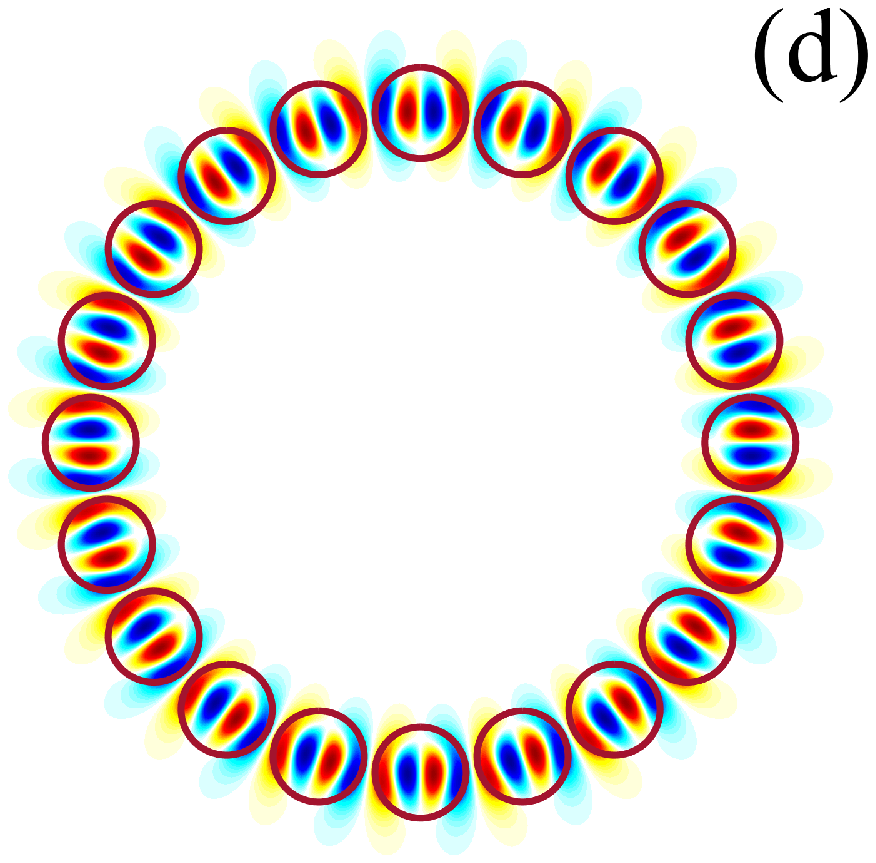}
\caption{Patterns of the symmetry protected BICs with $k_n=0$ in a
linear infinite array of rods (a) and (b) and their corresponding
counterparts in the circular array of 20 rods (c) and (d). The
parameters of BICs are the following: (a) $a=0.44, k_0=3.0758$,
(b) $a=0.44, k_0=3.5553$, (c) $a=0.44, k_0=3.0797, Q=4.8\cdot
10^7$, (d) $k_0=3.5461, Q=1.2\cdot 10^7$.} \label{fig6}
\end{figure}

Fig. \ref{fig5} (a) shows the pattern of electric field $E_z$ of
the symmetry protected BIC in the linear array of dielectric rods.
Fig. \ref{fig5} (b) and (c) show its circular counterparts in the
circular array of 15 and 25 rods respectively. Fig. \ref{fig6}
shows other examples of the symmetry protected BICs given in Refs.
\cite{PRA2014,Hu&Lu}.

\begin{figure}[ht]
\includegraphics[width=0.4\linewidth,clip=]{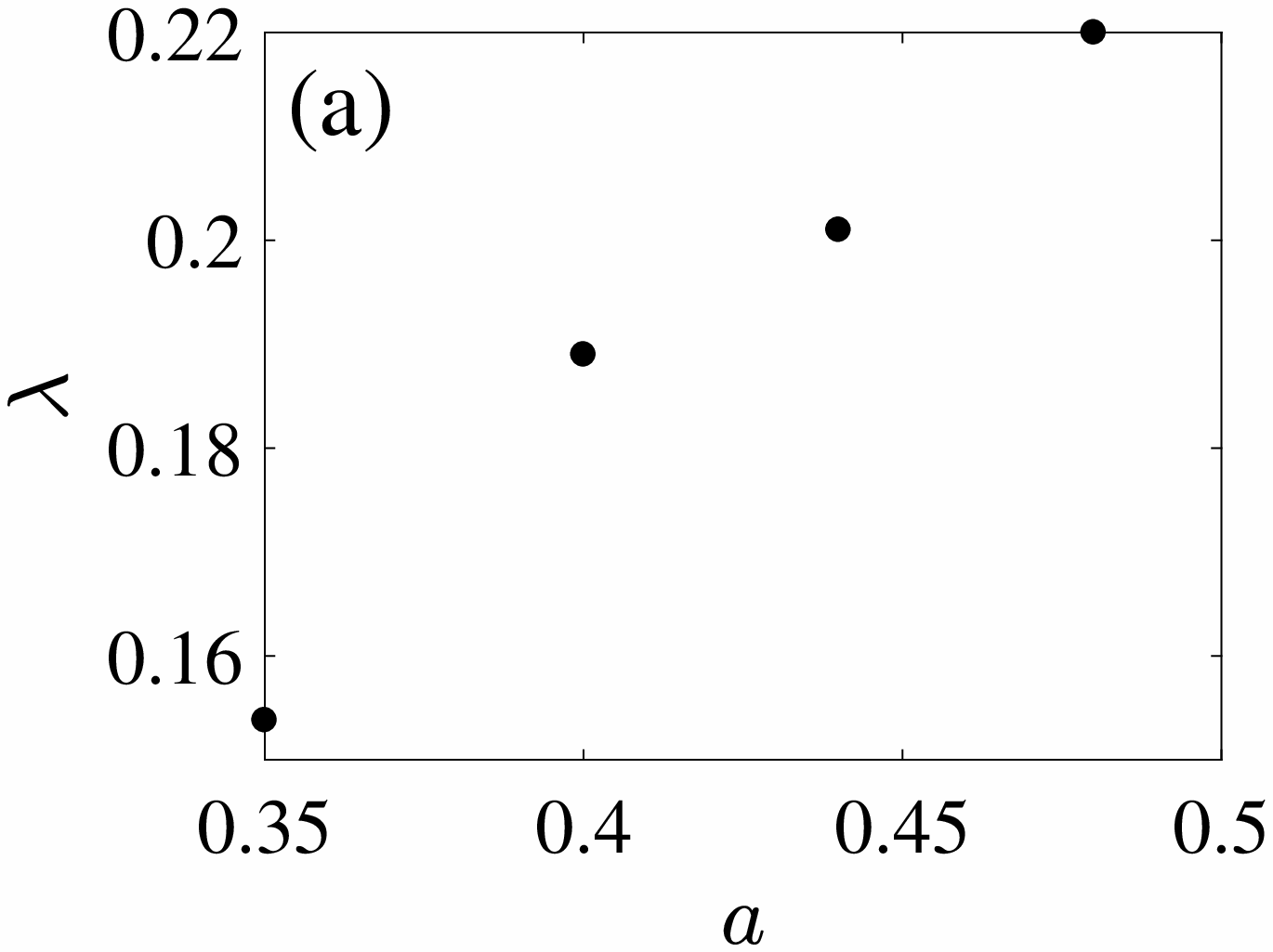}
\includegraphics[width=0.4\linewidth,clip=]{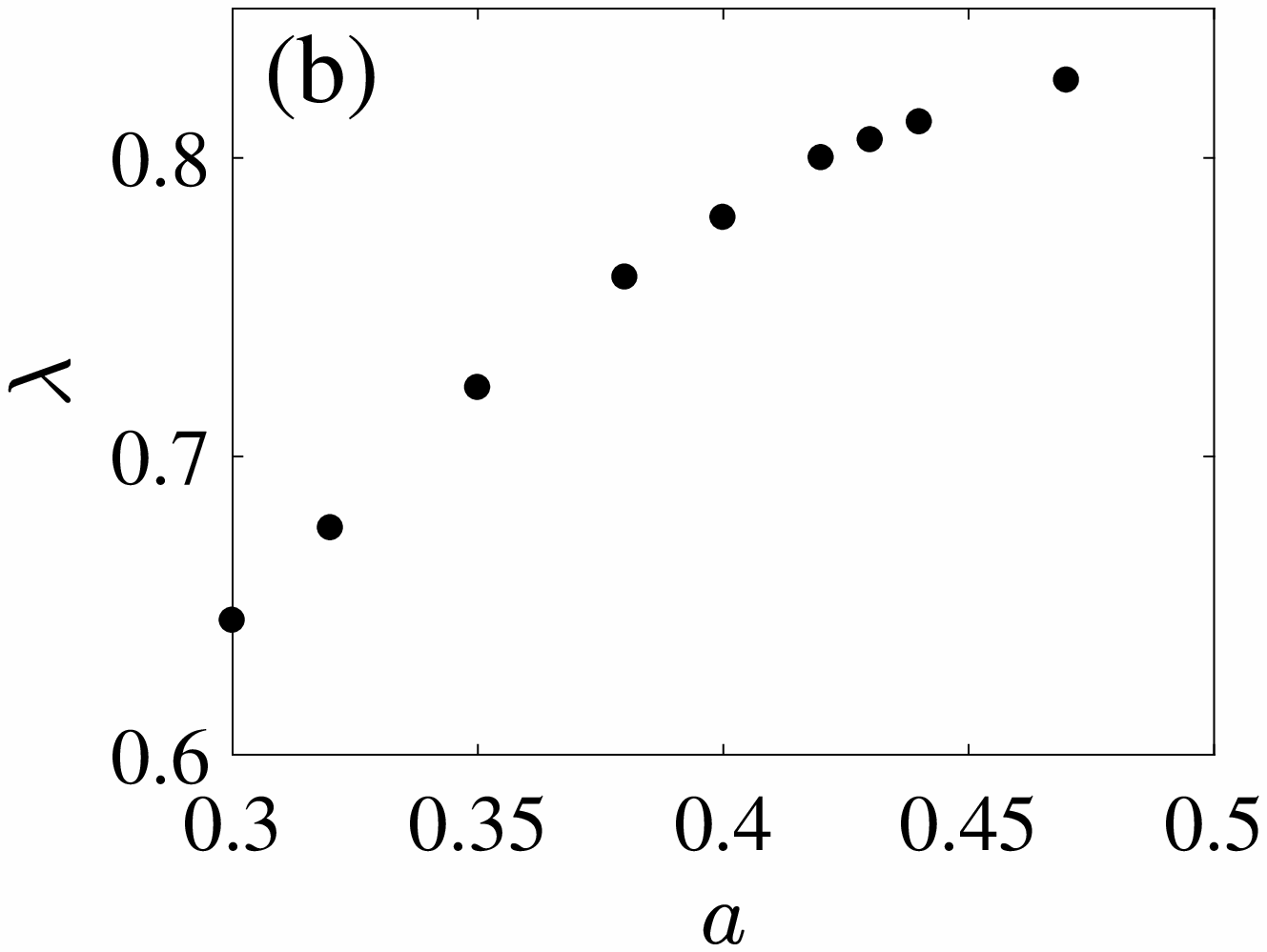}
\caption{Dependence of the parameter $\lambda$ which defines the
Q-factor $e^{\lambda N}$ vs the radius of rods for the symmetry
protected near-BICs at (a) $\epsilon=3$ and (b) $\epsilon=15$.}
\label{lambda}
\end{figure}

The patterns of the near-BICs as well as their exponentially large
Q-factors point out an analogy with the whispering gallery modes
(WGM) shown in Fig. \ref{fig5} (d) which also show exponentially
large Q-factor \cite{Oraevsky}. However in the present case the
Q-factor proportional to $Q\sim exp(\lambda N)$ while for the WGM
$Q\sim \exp(\kappa m)$ where $m$ is the order of the Bessel
function. Finally in Fig. \ref{lambda} we present numerical
results for the parameter $\lambda$ as dependent on the material
parameters.

\subsection{Non-symmetry protected BICs embedded into the first diffraction continuum}

\begin{figure}[h]
\includegraphics[width=0.3\linewidth,clip=]{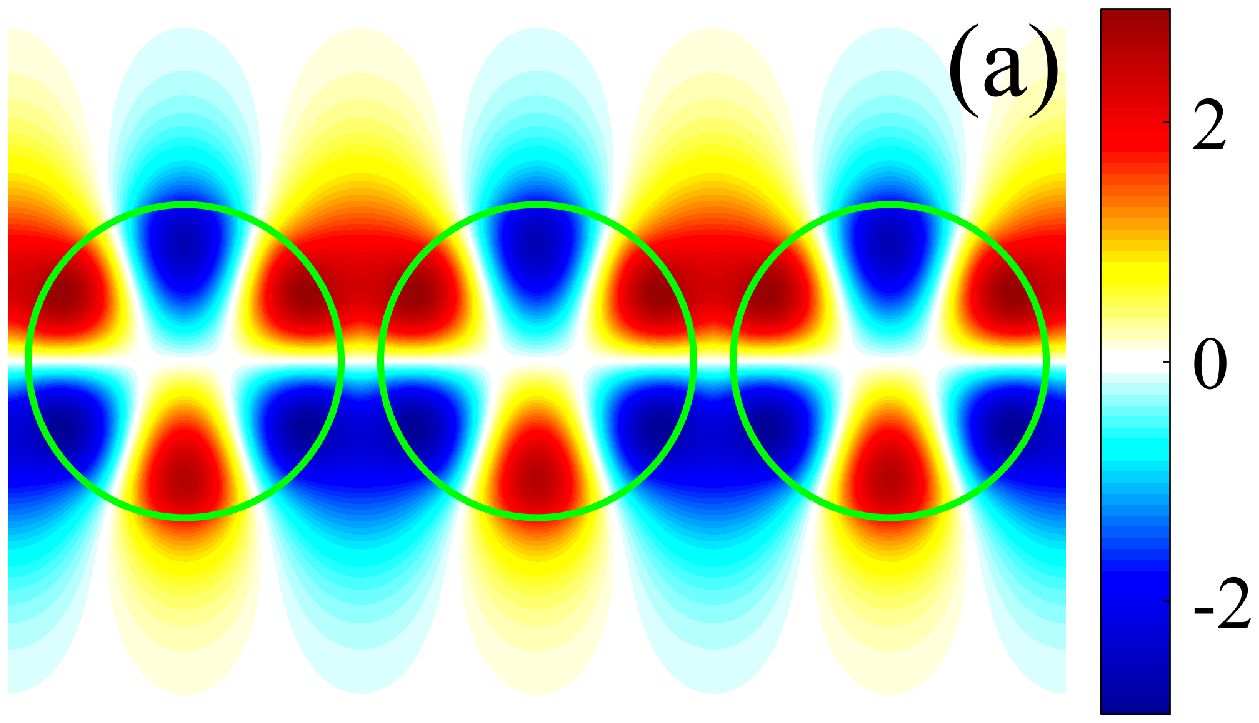}
\includegraphics[width=0.45\linewidth,clip=]{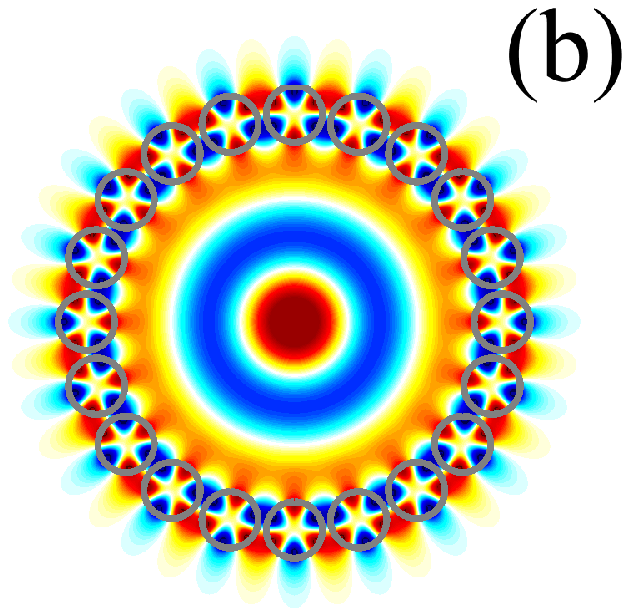}
\includegraphics[width=0.4\linewidth,clip=]{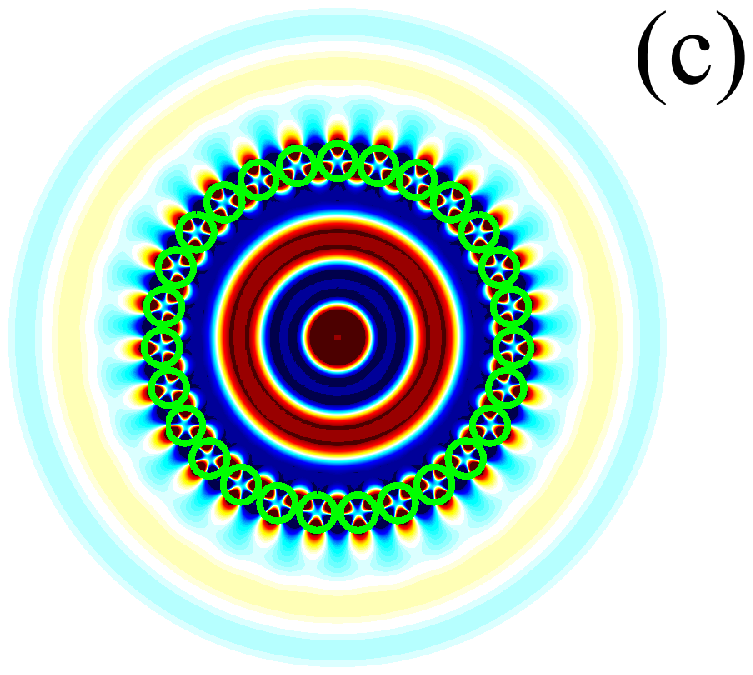}
\includegraphics[width=0.4\linewidth,clip=]{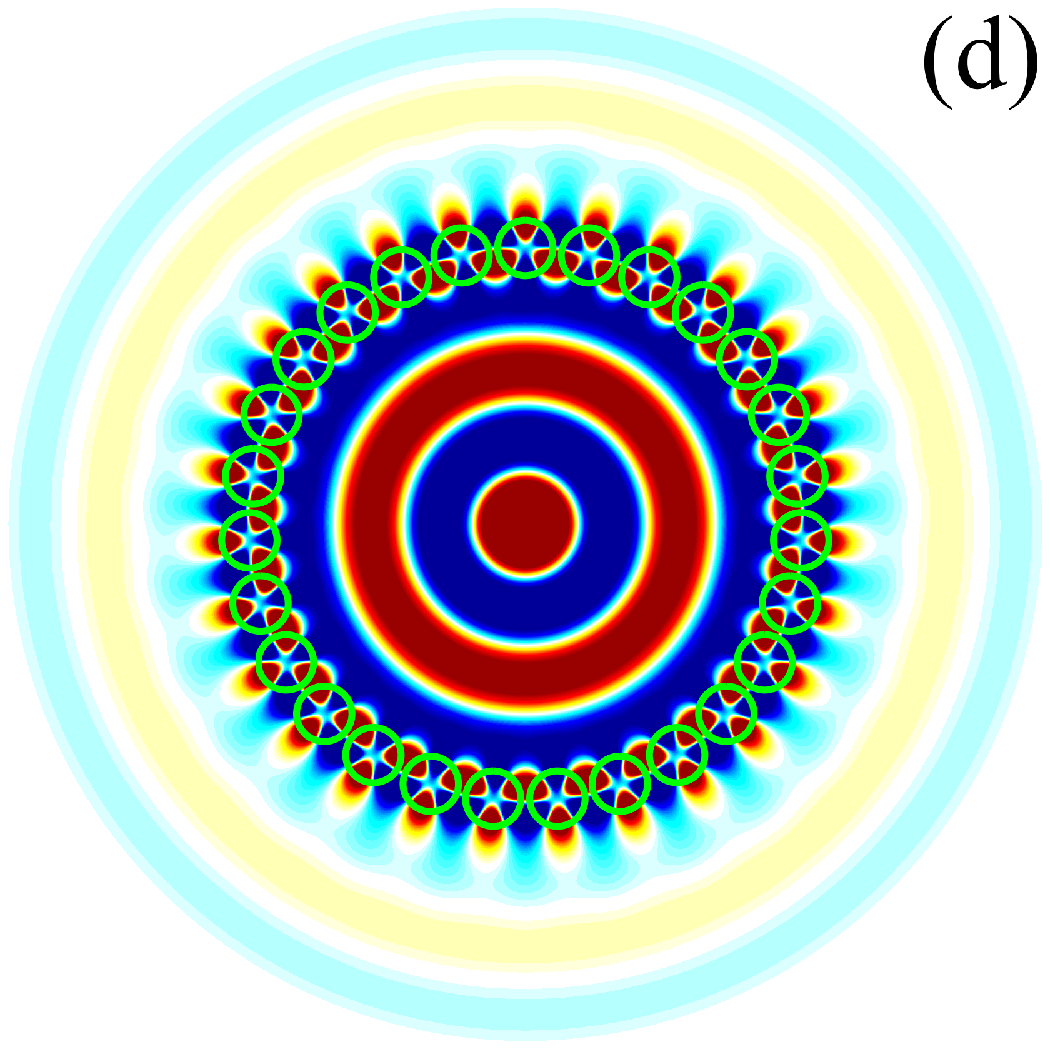}
\caption{Patterns of the non-symmetry protected BICs in a linear
infinite array of rods (a)  and their counterparts with $k_n=0$ in
the circular array of 20 (b), 10 (c) and 27 (d) rods. The
parameters of BICs are the following: (a) $a=0.44441, k_0=2.8299$,
(b) $a=0.43087, k_0=2.9234, Q=1.5\cdot 10^8$, (c) $a=0.43087,
k_0=2.9258, Q=150$, and (d) $a=0.43087, k_0=2.9251, Q=5000$.}
\label{fig7}
\end{figure}

However the above analogy of BICs in the circular array of rods
with the WGMs is terminated if to consider the non-symmetry
protected BICs which need tuning of the rod radius. Some examples
of these BICs borrowed from Refs. \cite{PRA2014,Yuan&Lu} are shown
in Fig. \ref{fig7} (a). Its counterpart in the circular array is
even for $k_n=0$ relative to $y'\rightarrow -y'$ in local system
of coordinates. In that case Eq. (\ref{BIC}) can be fulfilled by
tuning, for example, the rod radius $a$.  Figs. \ref{fig7} (c) and
(d) demonstrate what happens with these BICs if to forld the rods
in circle and then optimize the rod radius $a$. Because of the
symmetry of the infinite array of rods in respect to up and down
by tuning of the rod radius we can achieve zero coupling of the
BIC with both half radiation spaces above and below of the array.
In the circular array of rods we can achieve extremely small
coupling of the near-BIC to trap light against emanation outside
the circle. However we can not simultaneously suppress emanation
inside the circle. As the result one can see that the BIC mode
fills whole inner space of the circular array as demonstrated in
Fig. \ref{fig7}.

\begin{figure}[h]
\includegraphics[width=0.35\linewidth,clip=]{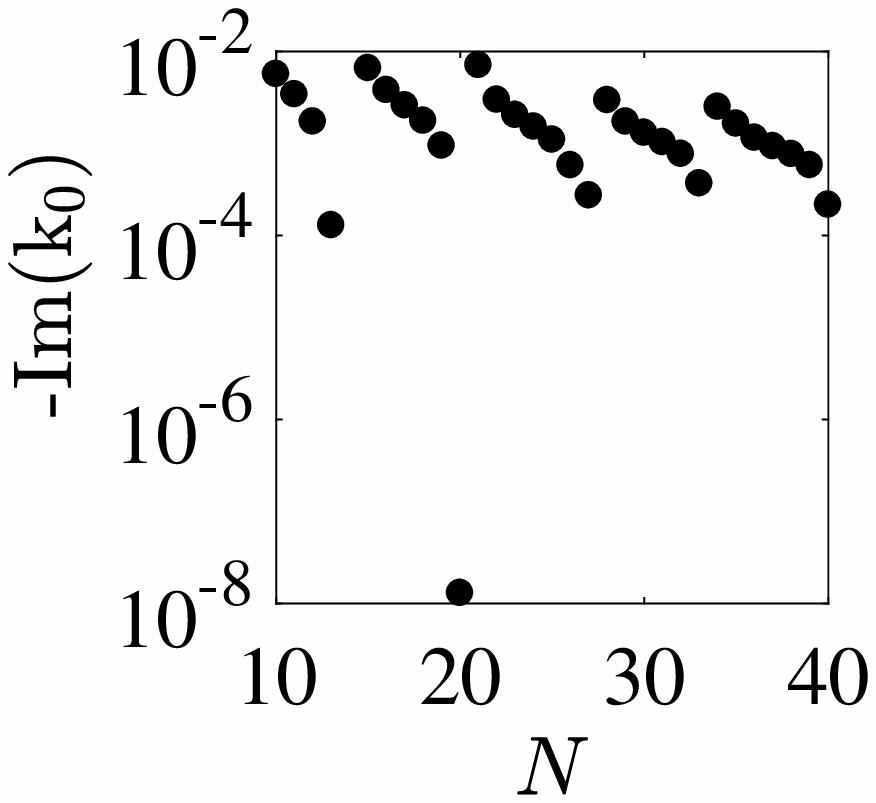}
\caption{The imaginary part of the frequency in log scale vs the number of rods for case of the
non-symmetry protected BIC shown above in Fig. \ref{fig7} (b) with
the rod radius $a=0.4387$ optimized for the case $N=20$.}
\label{fig8}
\end{figure}

The next aspect of the  non-symmetry protected near-BICs is related
to dependence of the Q-factor on $N$. For each $N$ the near-BIC
needs in optimization of the rod radius to give rise to extremely
large Q-factor similar to the symmetry protected near-BICs (see Fig.
\ref{fig4} (a)). However in practice it is easier to optimize the
rod radius for some selected number of rods. Currently we selected
$N=20$. Then change of the number of rods with the same radius $a$
gives the dependence of the Q-factor shown in Fig. \ref{fig8}
which has non-monotonic behavior. One can see that for
all $N$ except $N=20$ the circular array of rods can support only
resonances with the Q-factors in the range from hundreds till ten
thousands.
Let us leave the  material parameters unchange but take the number
of rods, say, $N\neq 20$. Then  the solution becomes resonant
state which strongly emanates EM field into the first diffraction
radiation continuum as shown in Fig. \ref{fig7} (c) and (d). The
quality factors are taken from Fig. \ref{fig8}.

\begin{figure}[ht]
\includegraphics[width=0.4\linewidth,clip=]{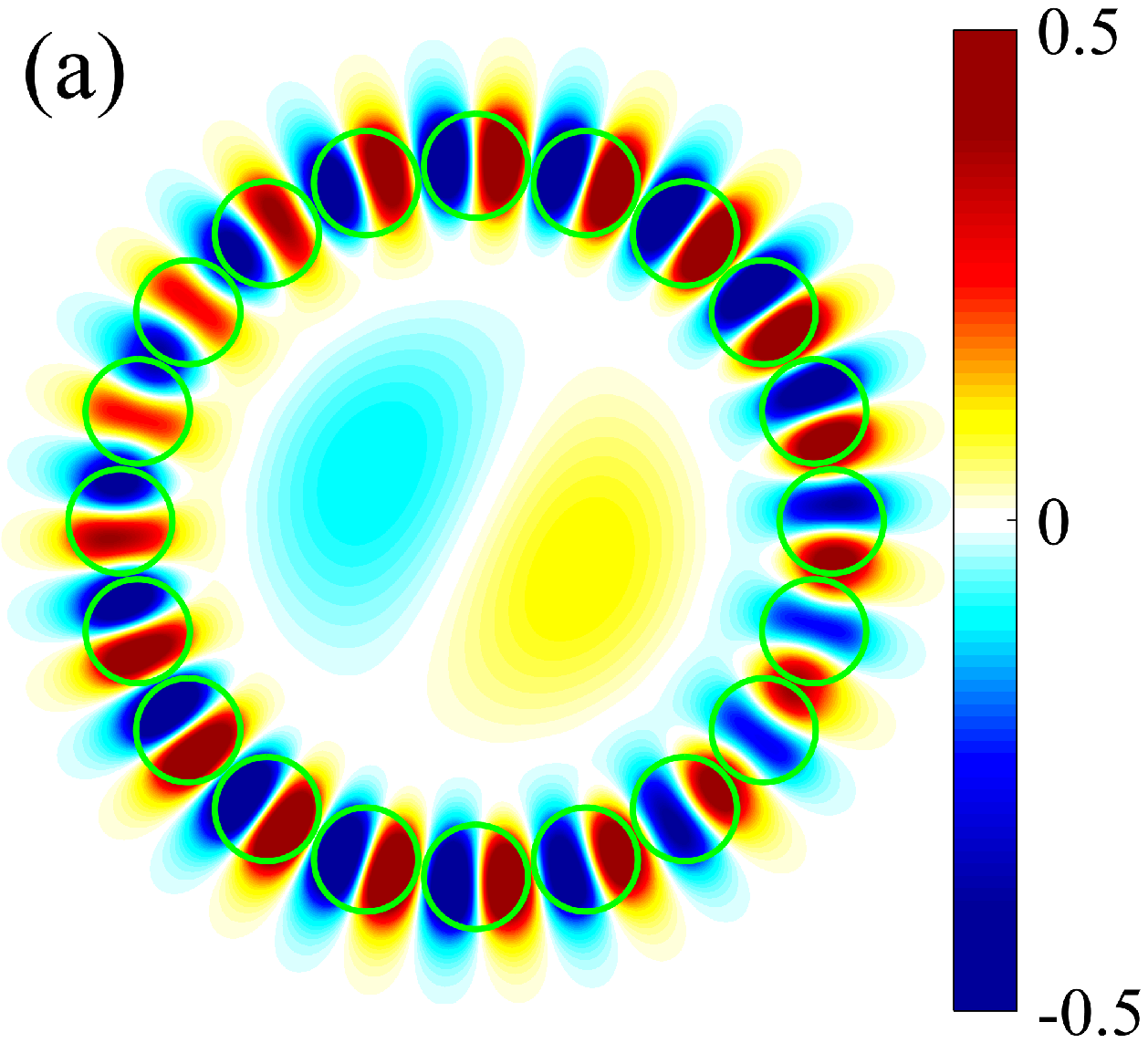}
\includegraphics[width=0.45\linewidth,clip=]{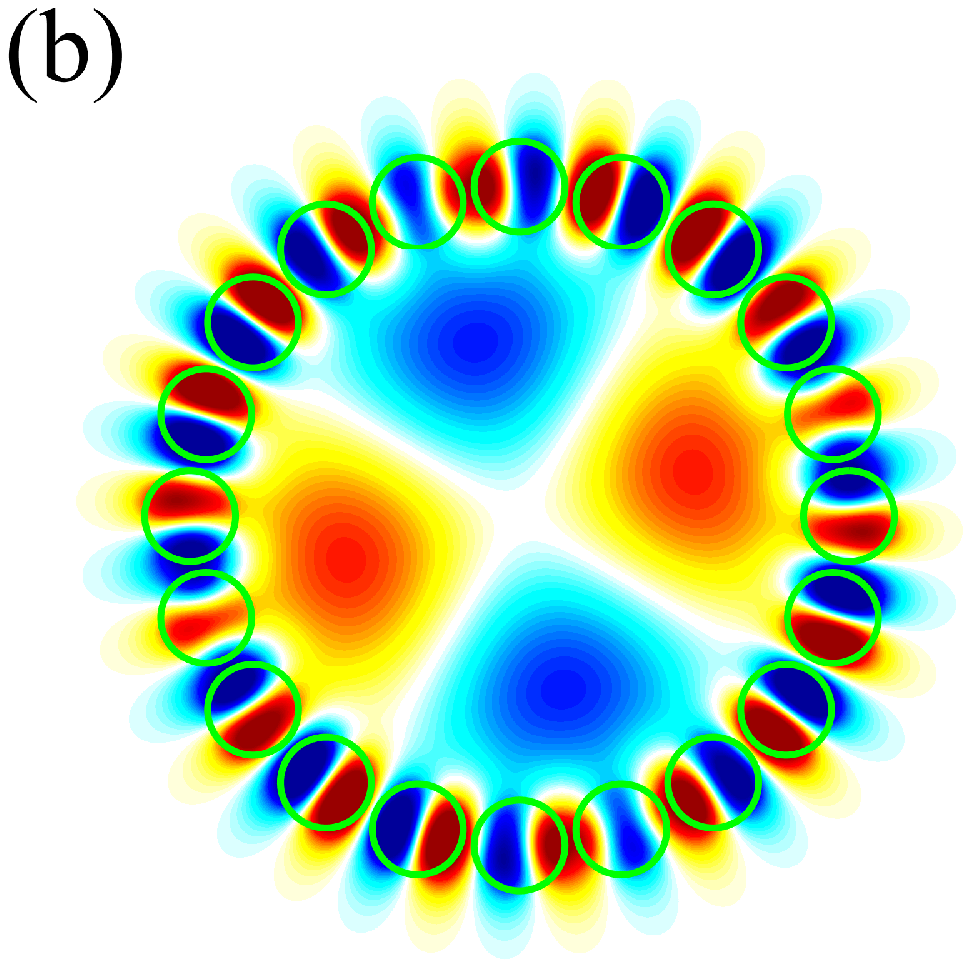}
\includegraphics[width=0.45\linewidth,clip=]{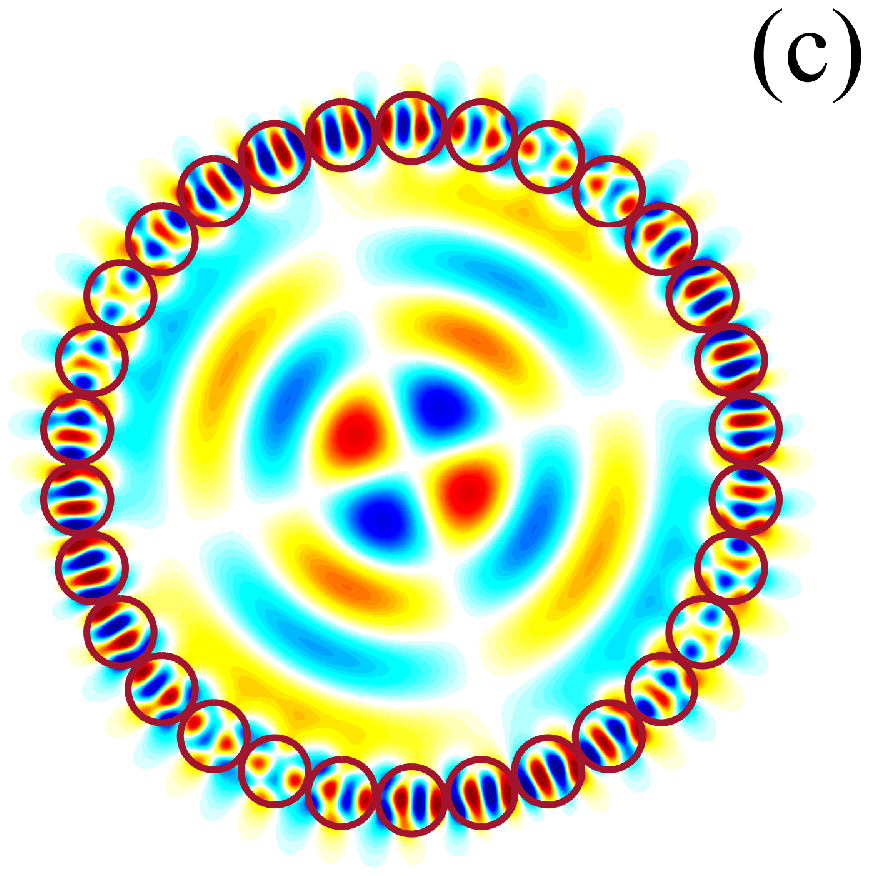}
\caption{Patterns of the non-symmetry protected Bloch BIC in
circular array of rods $N=20$. (a) $n=1, a=0.46636, k_0=1.7722,
Q=1.3\cdot 10^7$, (b) $n=2, a=0.4389, k_0=1.7663, Q=2.5\cdot 10^6$
and (c) $N=30, a=0.47745, k_0=3.39887, n=2, Q=5.3\cdot 10^8$.}
\label{fig9}
\end{figure}

Till now we considered near-BICs with the zero Bloch vector
$k_n=0$, i.e., with no angular dependence as shown in Fig.
\ref{fig7} (b). Fig. \ref{fig9} (a) shows the near-BIC with OAM
$n=1$ and Fig. \ref{fig9} (b), and (c) show the near-BICs with
$n=2$.
\begin{figure}[ht]
\includegraphics[width=0.35\linewidth,clip=]{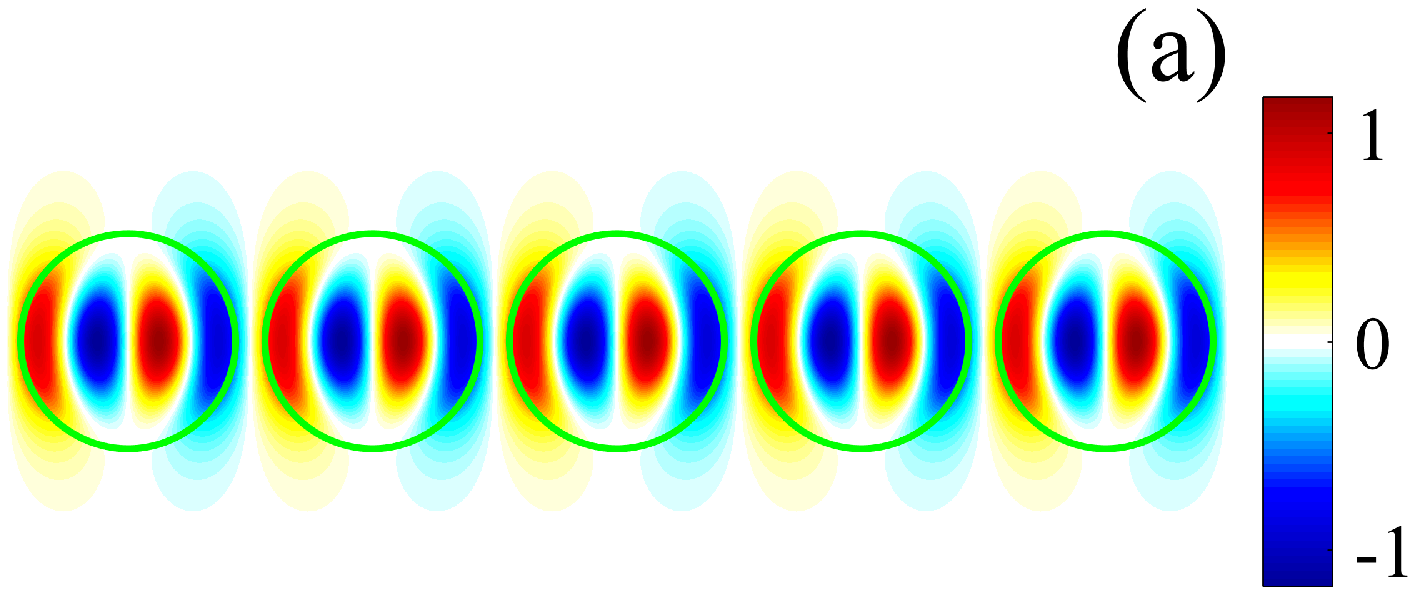}
\includegraphics[width=0.4\linewidth,clip=]{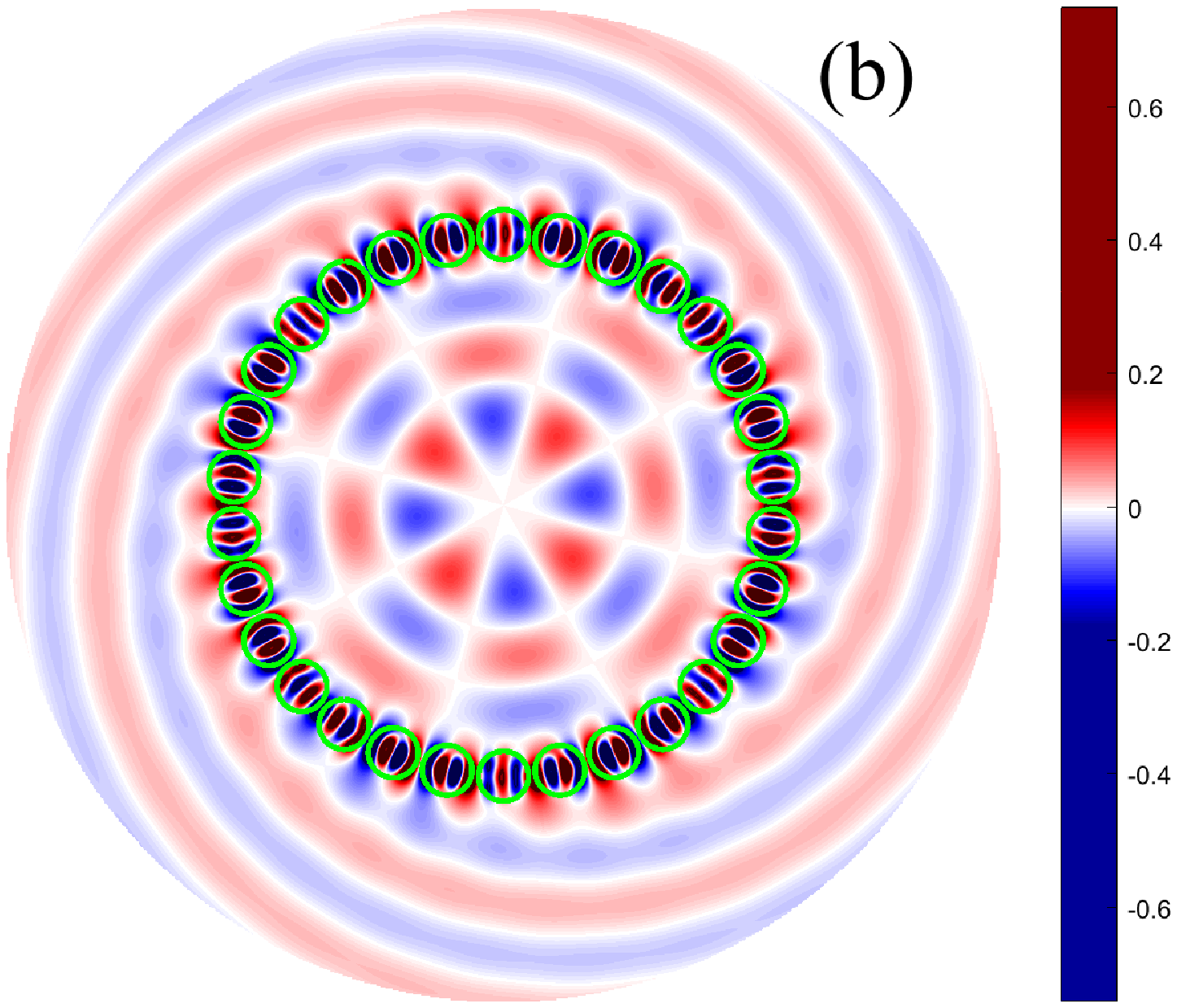}
\caption{Pattern of the BIC in the linear array with parameters
$a=0.44, k_0=3.5553$ (a) whose circular counterpart performs into
the radiating resonant mode with OAM with the parameters $a=0.44,
k_0=3.4469, n=4, Q=2.6\cdot 10^3 $.} \label{fig10}
\end{figure}
\subsection{BICs embedded into a few diffraction continua}
\begin{figure}[h]
\includegraphics[width=0.4\linewidth,clip=]{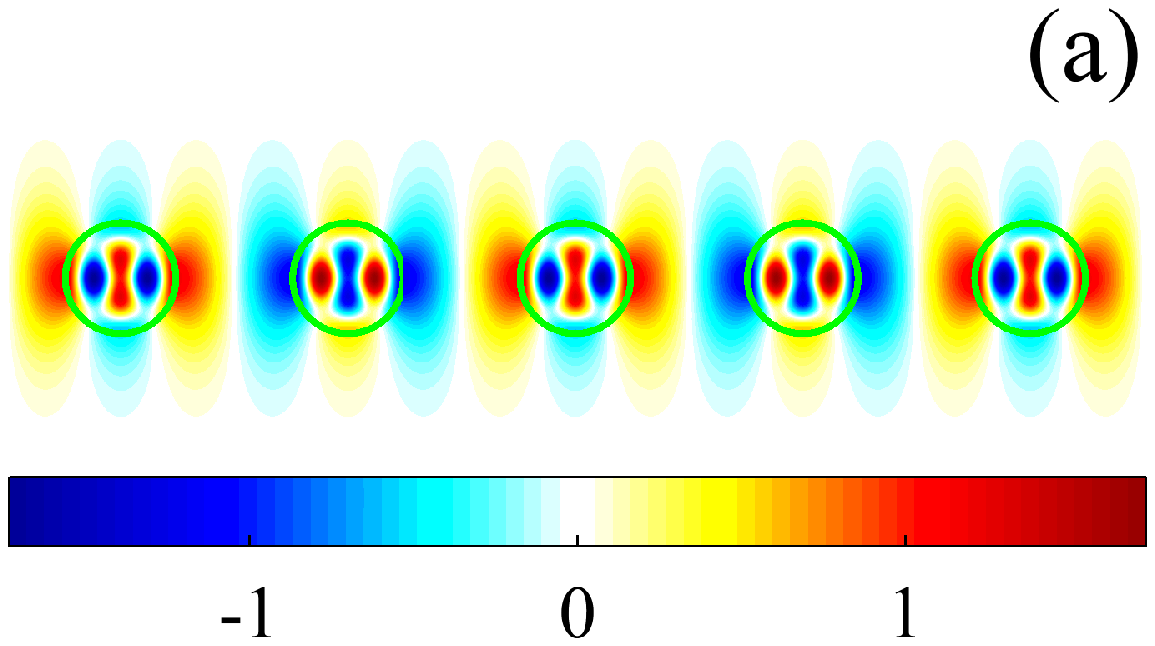}
\includegraphics[width=0.35\linewidth,clip=]{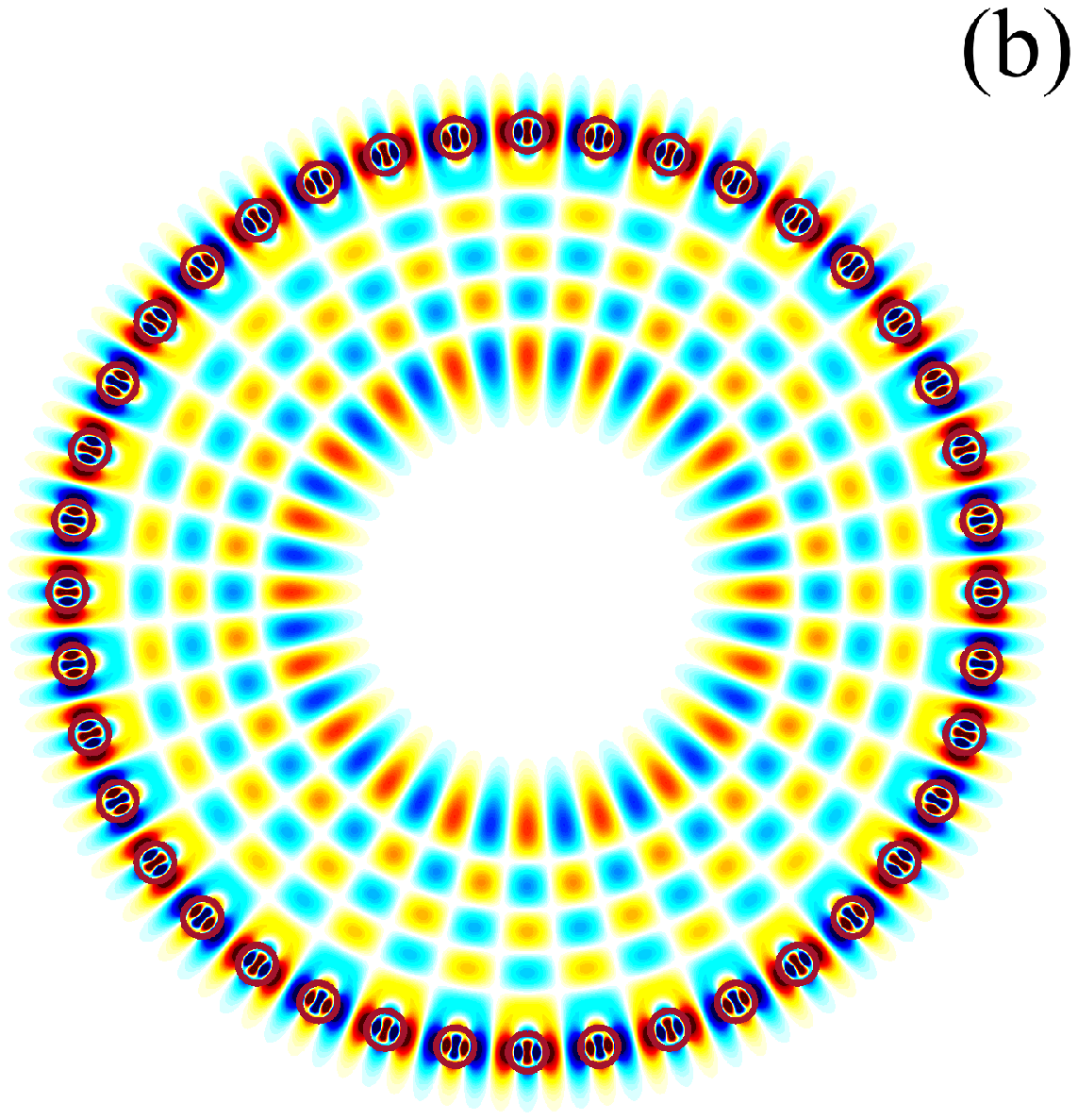}
\caption{(a) Pattern of the non-symmetry protected BIC embedded
into two diffraction continua in a linear infinite array of rods
and (b) their counterparts in the circular array of 40 rods. The
parameters of BICs are the following: (a) $a=0.24326, k_0=7.2946$,
(b) $a=0.2626, k_0=6.8092, Q=8.5\cdot 10^8$.} \label{fig11}
\end{figure}

In the infinite linear array of rods there are also BICs embedded
into a few diffraction continua given by Eqs. (\ref{psizn}) and (\ref{diffr cont}).
They are symmetry
protected relative to the first continuum and tuned by the radius of rods to be
embedded into the other continua. \cite{PRA2014}. Owing
to high frequencies these BICs occur at the rod radius smaller
compared to the BICs embedded into the first diffraction continuum
only. In this section we present their circular counterparts of such
BICs. We begin with the BIC embedded into the first and second
diffraction continua given by $p=0$ and $p=1$ of the linear array of rods for $\pi<k_0<3\pi$.
It  has the Bloch vector along the array equaled to $\beta=\pm \pi/h$
\cite{PRA2014} as shown in Fig. \ref{fig11} (a).
Respectively its circular counterpart has the same Bloch vector
$k_{N/2}=\pm \pi$.

\begin{figure}[h]
\includegraphics[width=0.35\linewidth,clip=]{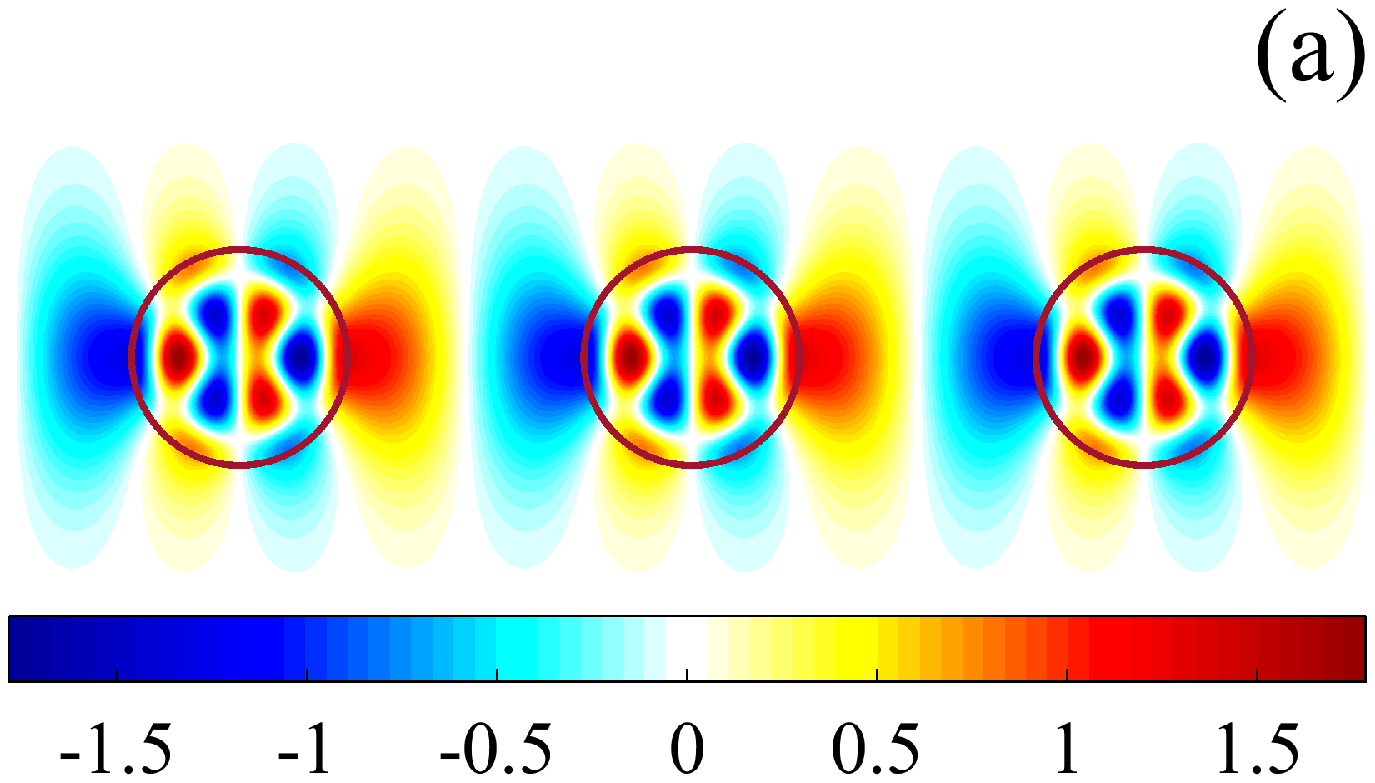}
\includegraphics[width=0.4\linewidth,clip=]{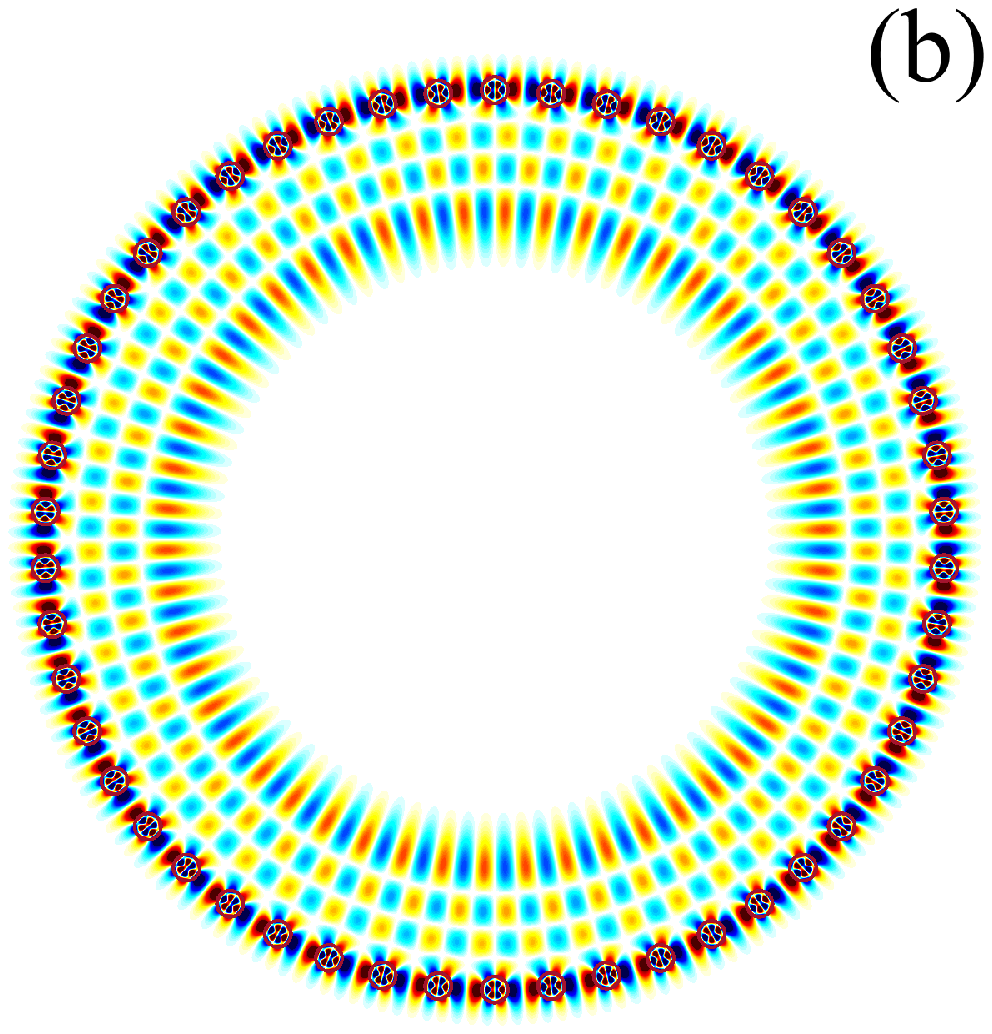}
\caption{(a) Pattern of the non-symmetry protected BIC embedded
into three diffraction continua in a linear infinite array of rods
and (b) their counterparts in the circular array of 50 rods. The
parameters of BICs are the following: (a) $a=0.2386, k_0=8.9518$,
(b) $a=0.2335, k_0=9.1298, Q=1.1\cdot 10^{14}$.} \label{fig12}
\end{figure}

Fig. \ref{fig12} (a) presents the BIC with the Bloch vector $\beta=0$
embedded into three diffraction continua with $p=0, p=\pm 1$
for $2\pi<k_0<4\pi$ \cite{PRA2014} and Fig. \ref{fig12} (b) shows its circular counterpart.
However in order to achieve
high Q-factors of these BICs the number of rods is to be rather high, 40 and 50.
\begin{figure}[h]
\includegraphics[width=0.4\linewidth,clip=]{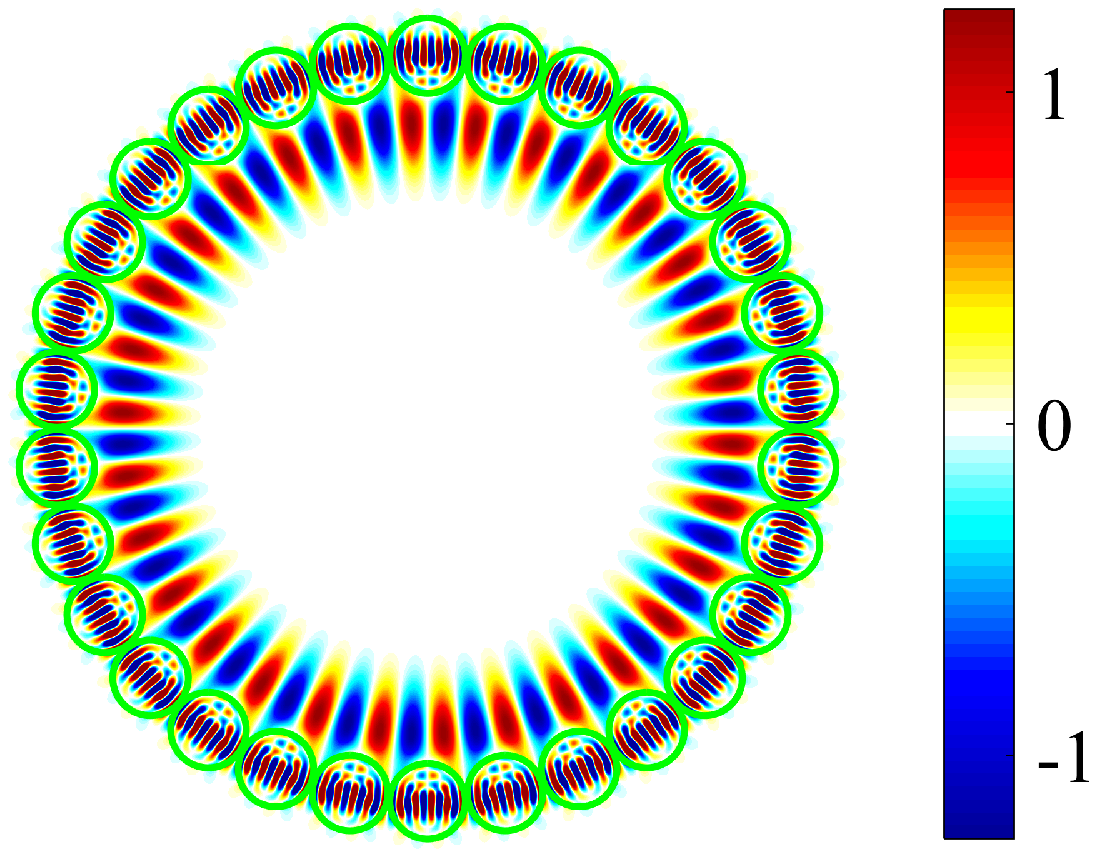}
\caption{Pattern of the near-BIC embedded into three diffraction continua
in the circular array of 30 rods with parameters:
$a=0.48365, k_0=8.3214, Q=1.852\cdot 10^{11}$.}
\label{fig13}
\end{figure}
Above we have presented the symmetry protected BICs which localized
around the rods (see Figs. \ref{fig5} and \ref{fig6}) and
non-symmetry protected BICs which fill whole inner space of the circular array
(see Figs. \ref{fig7}, \ref{fig9} and \ref{fig10}).
One can see that BICs embedded into two or more diffraction continua (see Figs. \ref{fig11}
and \ref{fig12}) have radial range of localization less than the radius of circle $R$.
That radial behavior of the BICs is results of radial behavior  of the Bessel functions
of high order. For $\nu\ll 1$ we have an asymptote through the Airy function \cite{Abramowitz}
$$J_{\nu}(\nu+z\nu^{1/3})\sim 2^{1/3}\nu^{-1/3}Ai(-2^{1/3}z)+O(\nu^{-1}),$$
The Airy function tends to zero when its argument exceeds zero. From Eq. (\ref{Ezinside2}) we have $\nu=n+qN$ and $z=k_0r$.
Therefore the radial width of BIC localization can be evaluated as
\begin{equation}\label{rad local}
\Delta=R-r\sim R-\frac{n+qN}{k_0}.
\end{equation}
 If the BIC is embedded into only the first diffraction continuum we have $k_0<2\pi$. Then from Eq.
 (\ref{rad local}) we obtain that the BIC occupies whole inner region inside the circle.
 For the BIC embedded
 into the first and second diffraction continua we have $\pi< k_0<3\pi$ that gives the
 $$\Delta\sim R-\frac{N}{2k_0}=N\left(\frac{1}{2\pi}-\frac{1}{2k_0}\right).$$
 In particular for the near-BIC shown in Fig. \ref{fig12} we have $n=N/2, q=0$ and $k_0=6.8$ to obtain
$\Delta\sim R(1-\pi/k_0)=0.54R$ that is close the numerical result shown in Fig. \ref{fig12}.
 At last, for the BIC embedded into three  continua we have $n=0, k_0=8.32$ and respectively from Eq. (\ref{rad local})
 obtain $$\Delta\sim R\left(\frac{1}{2\pi}-\frac{1}{k_0}\right)\approx \frac{R}{4}$$ that again is
 in good agreement with Fig. \ref{fig13}.
Surprisingly we revealed the near-BIC with $Q=1.85\cdot 10^{11}$
shown in Fig. \ref{fig13} in the circular array of 30 rods whose
linear counterpart is not the BIC but the narrow resonance.

\section{Summary}
We considered light trapping by circular array of infinitely long
dielectric rods. Each BIC, symmetry protected and non-symmetry
both, found in the linear arrow of rods
\cite{PRA2014,Hu&Lu,Yuan&Lu} has its circular counterpart,
near-BICs. Although the trapped light modes can not be rigorously
considered as the BIC in the circular array of rods because of
arguments presented in Refs. \cite{Colton,Silveirinha}, we have
presented analytical arguments in favor that the Q-factor of
the symmetry protected BICs with zero Bloch number grows exponentially with the number of rods.
Numerically this important result presented in Fig. \ref{fig4} (a)
and independently by Lu and Liu \cite{Lu&Liu}. In particular for
25 rods the Q-factor of the symmetry protected trapped modes
reaches values of order $10^{15}$ similar to the whispering gallery modes as demonstrated in
Fig. \ref{fig5}. In practice such Q-factors make the near-BICs in
the circular array indistinguishable from true BIC in the infinite
array of rods that allowed us to define them as the near-BICs.
The symmetry protected near-BICs with zero Bloch number are close in nature to the
whispering gallery modes (WGMs) (see Fig. \ref{fig5} (d)) whose
high Q factor is explained by total internal reflection but not by
destructive interference.

However the analogy with the WGM is ended if to proceed to
the symmetry protected near-BICs with non zero Bloch number or the non-symmetry protected near-BICs.
The cardinal difference between these near-BICs and the WGM is that the former
fills whole inner space of the circular array. As dependent on the Bloch vector
$k_n=2\pi n/N, n=0, 1, 2, \ldots$ the inner structure of the near-BIC defines the orbital angular momentum
(OAM) in respect to circular array and irrespectively to the solution inside the individual rod.
After abrupt change of the radius of the circle these near-BICs with OAM emanate in the surrounding space in the
form shown in Fig. \ref{fig10}. Also the Q-factor of the non-symmetry protected near-BICs can be reached
extremely large however we can not conclude that there is exponential behavior of the Q-factor
with the number of rods $N$ because of necessity to tune the radius of rods for each $N$.
As it was shown in Ref. \cite{PRA2014} there are BICs in the infinite linear array of rods
embedded into a few diffraction continua. In the present paper we have presented counterparts of
these BICs in the circular array of rods. They compose rather interesting feature which is
partial filling of the inner space of circular array.
Note one can consider the H polarized near-BICs in the circular array of dielectric rods
with similar results.

{\bf Acknowledgments}: We acknowledge discussions with  Andrey
Bogdanov and Dmittrii Maksimov. This  work  was partially
supported  by Ministry  of  Education  and  Science  of  Russian
Federation (State contract  N  3.1845.2017) and RFBR grants
16-02-00314 and 17-52-45072.

\end{document}